\documentclass[11pt,oneside,letterpaper]{article}
\usepackage{amssymb}
\usepackage{caption}
\usepackage{amsmath,bm}
\usepackage[dvips]{graphicx}
\usepackage{setspace}
\usepackage{fancyhdr}
\usepackage[dvipsnames]{xcolor}
\usepackage{ifpdf}
\usepackage{graphicx}
\usepackage{cite}
\usepackage{rotating}
\usepackage{comment}
\usepackage[margin=2.5cm]{geometry}
\usepackage[colorlinks=true, linkcolor  = blue,urlcolor=blue,citecolor=violet]{hyperref}
\usepackage{soul}
\usepackage{relsize}
\usepackage{tikz}
\usepackage{pgfplots}
\usepgfplotslibrary{polar}
\usetikzlibrary{trees}
\usetikzlibrary{decorations.pathmorphing}
\usetikzlibrary{decorations.markings}
\usetikzlibrary{decorations.markings}
\usepackage{float}
\usepackage{empheq}
\usepackage{bbold}
\usepackage{etoolbox}
\patchcmd{\thebibliography}{\section*{\refname}}{}{}{}

\newcommand{\bi}{\begin{itemize}}
\newcommand{\ei}{\end{itemize}}
\newcommand{\bea}{\begin{eqnarray}}
\newcommand{\eea}{\end{eqnarray}}
\newcommand{\be}{\begin{equation}}
\newcommand{\ee}{\end{equation}}

\newcommand{\dd}{\mathrm{d}}

\numberwithin{equation}{section}
\allowdisplaybreaks  % allow page breaks in displayed eqs

\begin{document}

%\vskip 5mm
%\begin{center}
%{\footnotesize{Department of Mathematics, King's College London, the Strand, London WC2R 2LS, UK}}
%\end{center}
%\vskip 5mm

%\setcounter{tocdepth}{2}

\onehalfspacing

\begin{center}

~
\vskip4mm
{{\huge { The two-sphere partition function from timelike Liouville theory at three-loop order
\quad 
 }
  }}
\vskip5mm

\vskip2mm

\vskip10mm

%{\color{black}D}ionysios Annino{\color{black}s}$^{{\tikz\penguin[hat=blue!40!black,scale=0.2];}}$\, \& ~Be{\color{black}atrix M}\"uhlmann$^{{\tikz\penguin[ scale=0.2];}}$ \\ 

Beatrix M\"uhlmann \\ 

\end{center}
\vskip4mm
\begin{center}
{
\footnotesize
{Department of Physics, McGill University, Montreal, QC H3A 2T8, Canada
}}
\end{center}
\begin{center}
{\textsf{\footnotesize{beatrix.muehlmann@mcgill.ca
}} } 
\end{center}
\vskip5mm

\vspace{4mm}
 
\vspace*{0.6cm}

%\end{center}
\vspace*{1.5cm}
\begin{abstract}
\noindent
%We discuss the semiclassical expansion of the two-sphere partition function in two-dimensional quantum gravity coupled to conformal matter with large positive central charge. We test a conjecture made in \cite{} relating the path integral derivation of the two-sphere partition function to the same but stemming from the DOZZ formula. Our testing ground is the semiclassical large positive matter central charge expansion up to four-loops. Besides reveiling marvellous cancellations of all the UV divergences we are also able to ......  this conjecture up to this order in the expansion. 

\end{abstract}

While the Euclidean two-dimensional gravitational path integral is in general highly fluctuating, it admits a semiclassical two-sphere saddle if coupled to a matter CFT with large and positive central charge. In Weyl gauge this gravity theory is known as timelike Liouville theory, and is conjectured to be a non-unitary two-dimensional CFT. 
We explore the semiclassical limit of timelike Liouville theory by calculating the two-sphere partition function from the perspective of the path integral to three-loop order, extending the work in 2106.01665. We also compare our result to the conjectured all-loop sphere partition function obtained from the DOZZ formula. Since the two-sphere is the geometry of Euclidean two-dimensional de Sitter space our discussion is tied to the conjecture of Gibbons-Hawking, according to which the dS entropy is encoded in the Euclidean gravitational path integral over compact manifolds.

%{\color{black}D}ionysios Annino{\color{black}s}$^{{\tikz\penguin[hat=blue!40!black,scale=0.2];}}$\, \& ~Be{\color{black}atrix M}\"uhlmann$^{{\tikz\penguin[ scale=0.2];}}$ \\ 

\newpage

\tableofcontents

\newpage

\section{Introduction}
Very little is known about de Sitter space at the quantum level. In particular, there is no S-matrix as for an asymptotically flat spacetime, or correlation functions as we encounter in the AdS/CFT dictionary (for a review see e.g. \cite{Witten:2001kn,Anninos:2012qw,Spradlin:2001pw}).
Because of the accelerated expansion, an observer in a de Sitter spacetime is surrounded by a cosmological horizon. Conjecturally a finite entropy is associated to this horizon, which in our Universe is of order $S_{\text{dS}}= 10^{120}$. 
Macroscopically, Gibbons and Hawking \cite{Gibbons:1976ue,Gibbons:1977mu} conjectured that the entropy of a de Sitter universe is encoded in the path integral 
\begin{equation}\label{eq:SdS}
e^{S_{\text{dS}}} =\sum_{\mathcal{M}} \int_{\mathcal{M}} [\mathcal{D}g]e^{-S_E[\Lambda,g_{ij},\mathcal{M}]}Z_{\text{matter}}[g_{ij},\mathcal{M},c_m]~,
\end{equation}
where $\Lambda >0$ is the cosmological constant and we are integrating over compact manifolds $\mathcal{M}$; $S_E$ is the Euclidean Einstein-Hilbert action with dominant sphere saddle. The round sphere is the geometry of Euclidean de Sitter space. The sphere is the analytic continuation of both the global and the static dS patch. We also include a matter CFT with central charge $c_m$. 

There are several questions about (\ref{eq:SdS}) which require further exploration. Firstly, no microscopic model for the de Sitter entropy is known. Moreover, performing the path integral (\ref{eq:SdS}) in general dimensions is a difficult task. Recent developments include \cite{longsphere,timelike, beatrix, Dio_beatrix,Law:2020cpj,David:2021wrw,Anninos:2021ihe,Hikida:2021ese}. 
In this paper, following the spirit of \cite{timelike, beatrix, Dio_beatrix, Anninos:2020geh,Anninos:2020ccj}, we restrict to a two-dimensional spacetime and explicitly calculate the gravitational path integral (\ref{eq:SdS}) in Weyl gauge \cite{David:1988hj,Distler:1988jt}. While the path integral of two-dimensional quantum gravity is in general highly fluctuating, if coupled to a matter CFT with large and positive central charge $c_m$, it admits a semiclassical two-sphere saddle. Vanishing conformal anomaly then implies that the Liouville central charge is large and negative, leading to timelike Liouville theory (TLT). TLT is conjectured to be a CFT \cite{Harlow:2011ny,Bautista:2019jau, timelike}.
Unlike in (spacelike) Liouville theory, in TLT \cite{Harlow:2011ny,Bautista:2019jau, timelike} the sign of the kinetic term is reversed. In Weyl gauge $g= e^{2\beta\varphi}\tilde{g}$ the action of TLT on a two-sphere is given by 

\begin{equation}\label{eq:StL}
S_{tL}[\varphi]= \frac{1}{4\pi} \int_{S^2} \dd^2x\sqrt{\tilde{g}} \left( -\tilde{g}^{ij} \partial_i \varphi \partial_j \varphi -  q \tilde{R} \varphi + 4\pi \Lambda e^{2\beta\varphi}  \right)~.
\end{equation}
In the above $\varphi$ denotes the Weyl mode, $\tilde{g}$ the fiducial metric with Ricci scalar $\tilde{R}$; $\Lambda > 0$ is the cosmological constant. Furthermore $q= \beta^{-1}-\beta$ and the timelike Liouville central charge is given by $c_{tL}= 1-6q^2$.  Restricting to genus zero, the path integral of interest is now
\begin{equation}\label{ZL}%^{(0)}
\mathcal{Z}_{tL}[\Lambda] =  \frac{1}{\text{vol}_{PSL(2,\mathbb{C})}} \times \int [\mathcal{D}\varphi ] e^{-S_{tL}[\varphi]}~.
\end{equation} 
We calculate three-loop corrections on top of the two-sphere saddle, extending the work in \cite{timelike}. This allows us to explore the Gibbons-Hawking conjecture and could be useful to constrain a possible microscopic theory of a two-dimensional de Sitter universe. 

%the Gibbons-Hawking conjecture in a two-dimensional de Sitter universe, trying to find as much structure as possible. Any putative microscopic theory would need to capture and reflect this structure. 
%
%In two dimensions the sum over $\mathcal{M}$ in (\ref{eq:SdS}) is replaced by a sum over genera, since the compact manifolds $\mathcal{M}$ are compact Riemann surfaces of genus $h$. Following the conjecture of Distler-Kawai and David \cite{Distler:1988jt,David:1988hj} the path integral of two-dimensional quantum gravity in Weyl gauge is weighted by the Liouville action. While the path integral of two-dimensional quantum gravity is in general highly fluctuating, if coupled to a matter CFT with large and positive central charge $c_m$, it admits a semiclassical two-sphere saddle. Vanishing conformal anomaly then implies that the Liouville central charge is large and negative, leading to TLT. Thus TLT provides an arena to explore the Gibbons and Hawking proposal (\ref{eq:SdS}) in two spacetime dimensions.  \newline\newline
%\textbf{Timelike Liouville theory $\&$ analytically continued DOZZ formula.}
%Whereas two-dimensional quantum gravity is a widely explored arena, receiving renewed interest over the last couple of years, TLT has been less discussed. Some work on TLT includes the conformal bootstrap program, e.g. \cite{Ribault:2015sxa,Ikhlef:2015eua,Bautista:2019jau}. Another direction arises since TLT can be coupled to a matter theory with $c_m\geq 25$ providing a time dependent target space of a more cosmological nature \cite{Polchinski:1989fn,Cooper:1991vg}.

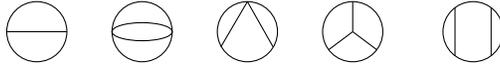
\begin{figure}[H]
\begin{center}
\begin{tikzpicture}[scale=.4]

\draw (-2.5,1) circle (1cm);
\draw[black] (-3.5,1) --(-1.5,1);

\draw (1,1) circle (1cm);
\draw (1,1) ellipse (1cm and .3cm);

\draw (4.5,1) circle (1cm);
\draw[black] (4.5,2) --(3.62,.5);
\draw[black] (4.5,2) --(5.35,.5);

\draw (8,1) circle (1cm);
\draw[black] (8,1) --(8,2);
\draw[black] (8,1) --(8.8,.4);
\draw[black] (8,1) --(7.2,.4);

\draw (12.,1) circle (1cm);
\draw[black] (12.6,1.8) --(12.6,.2);
\draw[black] (11.4,1.8) --(11.4,.2);

\end{tikzpicture}

\end{center}
\caption{Type of diagrams surviving at order $\mathcal{O}(\beta^4)$.}
\label{fig:diagramsBeta4_intro}
\end{figure}

Whereas the gravitational path integral (\ref{ZL}) admits a semiclassical, $\beta \rightarrow 0$, two-sphere saddle on top of which we can calculate loop corrections, there also exists a 
path integral independent approach toward the two-sphere partition function. Originally introduced for spacelike Liouville theory by Dorn-Otto \cite{Dorn:1994xn} and Zamolodchikov-Zamolodchikov \cite{Zamolodchikov:1995aa}, the DOZZ formula captures the three-point function of the Liouville vertex operators. The DOZZ formula has been extended to TLT by Zamolodchikov and Kostov-Petkova \cite{Zamolodchikov:2005fy,Kostov:2005av,Kostov:2006zp,Kostov:2005kk} (see also \cite{Harlow:2011ny}).  Since however e.g. the timelike DOZZ formula in \cite{Harlow:2011ny} for three area operators $\mathcal{O}_\beta = e^{2\beta\varphi}$ vanishes, contradicting the fact that the path integral yields a non-vanishing result, our comparison relies instead on analytically continuing the spacelike DOZZ formula, i.e. 
\begin{equation}
\langle \mathcal{O}_{\beta}(z_1) \mathcal{O}_{\beta}(z_2) \mathcal{O}_{\beta}(z_3) \rangle =  \frac{1}{\text{vol}_{PSL(2,\mathbb{C})}} \times \frac{C(b,b,b; \Lambda)|_{b\rightarrow \pm i\beta}}{|z_1-z_2|^2 |z_1-z_3|^2 |z_2-z_3|^2}~,
\end{equation}
where we also highlight the $\Lambda$ dependence of the structure constant. Below we summarise the sphere partition functions obtained from the two approaches described above. 
\newline\newline
\textbf{Results.}
%In this limit the path integral admits a round two-sphere saddle with all higher genus geometries suppressed.
\begin{itemize}
\item
Analytically continuing $C(b,b,b; \Lambda)|_{b\rightarrow \pm i\beta}$ and thrice integrating with respect to $\Lambda$ leads to the sphere partition function \cite{timelike,Giribet:2011zx}
\begin{align}\label{eq:sphere_DOZZ_intro}
&\mathcal{Z}^{\text{DOZZ}}_{tL}[\Lambda] %= -2\int^\Lambda \dd\Lambda' \int^{\Lambda'} \dd\Lambda'' \int^{\Lambda ''}\dd\Lambda''' C(b,b,b)|_{b\rightarrow \pm i\beta}\cr
=\pm i\left(\pi \Lambda \gamma(-\beta^2)\right)^{-\frac{1}{\beta^2}+1}\frac{(1+\beta^2)}{\pi^3 q \gamma(-\beta^2)\gamma(-\beta^{-2})}\,e^{q^2 -q^2 \log 4}\cr
 &\approx \pm  e^{-\frac{1}{\beta^2}- \frac{1}{\beta^2}\log (4\pi \beta^2) }\Lambda^{-\frac{1}{\beta^2}+1} \left( 1 - e^{\frac{2i\pi}{\beta^2}}\right)\cr
 &\times \left(\frac{1}{\beta}+ {\frac{1}{6}\left({19}- 6 \log 4\right)}\beta+ \left(\frac{1}{2}\times \frac{1}{36} (19-6 \log 4)^2-\frac{2}{3} \zeta (3)\right)\beta^3+... \right)~,
\end{align}
where $\gamma(x) \equiv \Gamma(x)/\Gamma(1-x)$.
This is conjectured to be the two-sphere partition function of TLT \cite{Giribet:2011zx}.
\item On the other side, evaluating (\ref{ZL}) on the round two-sphere saddle and calculating three-loop corrections on top of this saddle yields
%\begin{multline}\label{Ztl_oneLoop_noDiagrams}
%\mathcal{Z}_{tL}[\Lambda] \approx
%% \approx  \pm i\frac{a_0}{\text{vol}_{SO(3)}} \times e^{-\frac{1}{\beta^2}- \frac{1}{\beta^2}\log\left(4\pi  \beta^2\right)}\times \beta^2 \left(1+\frac{\beta^2}{2}+ \ldots\right)\times\left(\frac{1}{\beta^3}- \frac{3}{\beta}+\ldots\right)\cr
%%\times \left(1+\left(\frac{37}{12}- \log (16\pi)\right)\beta^2+ \ldots \right)\left(1-\left(\frac{3}{4}+ \frac{5\pi}{2}\frac{a_1}{a_0}\right)\beta^2+ \ldots \right) (\upsilon\Lambda_{\mathrm{uv}})^{\frac{1}{2}+ \frac{3}{2}+ \frac{5}{2}-\frac{10}{3}- \beta^2}\times (\upsilon\Lambda)^{-\frac{1}{\beta^2}+1}~\cr
% \frac{\pm i }{\text{vol}_{SO(3)}} \, \mathrm{const}\,\times \,e^{-\frac{1}{\beta^2}- \frac{1}{\beta^2}\log\left(4\pi  \beta^2\right)} \upsilon^{\frac{c_L}{6}}\Lambda_{\mathrm{uv}}^{\frac{7}{6}- \beta^2} \Lambda^{-\frac{1}{\beta^2}+1} \times \Big(\frac{1}{\beta} \cr
% - \left(\frac{1}{6}- \frac{5\pi}{2}\left(\frac{a_1}{a_0}\right)+2\gamma_E+ \log 4\pi - \mathlarger{\mathlarger{\ominus}}\right)\beta +()\beta^3+ \ldots  \Big)~,
%\end{multline}
\begin{align}\nonumber\label{Ztl_oneLoop_intro}
&~\mathcal{Z}_{tL}[\Lambda] \approx
{\pm i }e^{-\frac{1}{\beta^2}- \frac{1}{\beta^2}\log\left(4\pi  \beta^2\right)} \upsilon^{\frac{c_{tL}}{6}}\Lambda_{\mathrm{uv}}^{\frac{7}{6}- \beta^2} \Lambda^{-\frac{1}{\beta^2}+1} \times \Bigg(\frac{1}{\beta} + \bigg(\frac{1}{6}(19-6\log 4)- (2\gamma_E+\log \pi)\\ \nonumber
&+ \mathlarger{\mathlarger{\ominus}}- \frac{10}{3}+ \frac{5\pi a_1}{2a_0}\bigg)\beta+\bigg(\frac{1}{2}\times \left(\frac{1}{6}(19-6\log 4)- (2\gamma_E+\log \pi)+ \mathlarger{\mathlarger{\ominus}}- \frac{10}{3}+ \frac{5\pi a_1}{2a_0}\right)^2 \\ 
&+\text{loops}_{\beta^4}-\frac{17}{27}+ \frac{15\pi a_1}{4a_0}-\frac{25\pi^2a_1^2}{8a_0^2}-\frac{1}{2}\mathlarger{\mathlarger{\ominus}}^2\bigg)\beta^3+...\Bigg)~,
%& - \left(\frac{1}{6}- \frac{5\pi}{2}\left(\frac{a_1}{a_0}\right)+2\gamma_E+ \log 4\pi-  \mathlarger{\mathlarger{\ominus}} \right)\beta 
% +\bigg(\frac{1}{2}\left(\frac{1}{6}- \frac{5\pi}{2}\left(\frac{a_1}{a_0}\right)+2\gamma_E+ \log 4\pi-  \mathlarger{\mathlarger{\ominus}} \right)^2 \cr
%&+\frac{10\pi}{3}\left(\frac{a_1}{a_0}\right)- \frac{17}{27}
% -\frac{1}{2}\mathlarger{\mathlarger{\ominus}}^2 +\left(\frac{1}{6}- \frac{5\pi}{2}\left(\frac{a_1}{a_0}\right)+2\gamma_E+ \log 4\pi\right)\mathlarger{\mathlarger{\ominus}}+ 3\text{-loop}\bigg)\beta^3+ ... \Big)~,
\end{align}
where $a_1= 27a_0/(20\pi) $, $\gamma_E$ is the Euler-Mascheroni constant; $\Lambda_{\text{uv}}$ is the UV cutoff of our theory and $4\pi \upsilon$ denotes the area of the two-sphere. Finally $\ominus$ denotes a ``melonic'' type of diagram, whereas ``loops$_{\beta^4}$'' comprises all three-loop diagrams. These diagrams appear at order $\mathcal{O}(\beta^4)$, and after taking into considerations all cancellations are diagrams of the form shown in figure \ref{fig:diagramsBeta4_intro}. The gauge fixing of vol$_{PSL(2,\mathbb{C})}$ and the saddle create the expansion in odd powers of $\beta$ in (\ref{Ztl_oneLoop_intro}); 
\end{itemize}
It is immediately visible that $\mathcal{Z}^{\text{DOZZ}}_{tL}$ and $\mathcal{Z}_{tL}$ do not agree with each other. From the path integral perspective, the conformal anomaly of the sphere, $\upsilon^{c_{tL}/6}$, is immediate, and the UV cutoff $\Lambda_{\text{uv}}$ combines with the area and the cosmological constant to render the sphere partition function dimensionless. These dimensions must be reverse-engineered in (\ref{eq:sphere_DOZZ_intro}). Furthermore, whereas the forms of the small $\beta$ expansions agree with each other, the coefficients do not. 
This was noted already in \cite{timelike} and a conjectured solution was put forward. In this paper, by extending the loop corrections to third order, we could test this proposal. Taking scheme dependency into consideration and allowing the rescaling $\Lambda_{\text{uv}}\rightarrow s\Lambda_{\text{uv}}$, we conjecture that for $\log s= {-\left(2\gamma_E+\log\pi\right)}$ the small $\beta$ expansions of (\ref{eq:sphere_DOZZ_intro}) and (\ref{Ztl_oneLoop_intro}) agree not just to second but up to third order. Furthermore we observe a systematic cancellation of UV divergent diagrams as well as cancellations between UV finite diagrams. After the dust settles only the type of diagrams shown below survives at order $\mathcal{O}(\beta^4)$.
\noindent
These diagrams suggest a nice generalisation of the melonic type diagrams appearing at order $\mathcal{O}(\beta^2)$ to higher orders. Furthermore, we note that the agreement between(\ref{eq:sphere_DOZZ_intro}) and (\ref{Ztl_oneLoop_intro})  also at three-loop order provides further evidence that the semiclassical expansion leads to a loophole around Gribov-phenomena \cite{Gribov:1977wm}. 

A major difference between (\ref{eq:sphere_DOZZ_intro}) and (\ref{Ztl_oneLoop_intro})  clearly remains. This is the appearance of the term $e^{\frac{2i\pi}{\beta^2}}$ in $\mathcal{Z}^{\text{DOZZ}}_{tL}[\Lambda] $. From a path integral perspective, it can be interpreted as the contribution of a second \textit{complex} saddle. As we will explain in section \ref{sec:4}, in the spirit of \cite{Witten:2021nzp}, both saddles are allowed from a path integral perspective. 
\newline\newline
\textbf{Outline.} The outline of this paper is as follows. In section \ref{sec:2} we introduce TLT and explain its main features. We explain the round two-sphere saddle and the small fluctuations thereof. For details of the calculations we refer to \cite{timelike}. In section \ref{sec:3} we delve into the diagramatics. After introducing the propagator on the two-sphere we recap the two-loop diagramatics explored in \cite{timelike}. Our main calculations are the three-loop contributions studied in section \ref{sec:33}. In section \ref{sec:spherePI} we summarise the two- and three-loop contributions, explain the cancellations of the UV divergences, and present the TLT sphere partition function. In section \ref{sec:4} we compare our result for the two-sphere partition function to the one obtained upon analytically continuing the spacelike DOZZ formula for three area operators. We discuss scheme dependency and the allowability of the complex saddle. Finally section \ref{sec:5} provides some concluding and speculative remarks.
%Another way to obtain the two-sphere partition function is based on the path integral of two-dimensional quantum gravity. In the semiclassical limit $\beta\rightarrow 0^+$ this allows a sphere saddle. The two-sphere partition function can then be obtained by calculating fluctuations around this saddle. 
%This involves a lot of subtleties, not only do we need to take care of the infinite symmetry group of the two-sphere, namely the Moebius group, but we also need to fix the measure over the space of fields. The latter relies heavily on the ultraviolet cutoff. On the other side the UV cutoff has been set to one in the DOZZ formula. Matching the expressions therefore relies on a a posteriori definition of the UV cutoff in the DOZZ formula. 
%In \cite{timelike} we performed a two-loop analysis of the path integral of two-dimensional quantum gravity. We could match the linear $\mathcal{O}(\beta)$ term in (\ref{eq:sphere_DOZZ}) upon introducing a shift in the UV cutoff. In this note we would like to test this conjecture by calculating the three-loop contribution from the path integral perspective. 
\section{Timelike Liouville theory}\label{sec:2}
In this section we introduce the components of (\ref{ZL}). We will not delve into any details, for which we refer to \cite{timelike}.

We fix the background fiducial metric $\dd \tilde{s}^2$ to be the Fubini-Study metric on the two-sphere with area $4\pi\upsilon$. If we denote by $\varphi$ the Weyl mode we have 
\begin{equation}\label{metric_start}
\dd s^2= e^{2\beta\varphi} \dd \tilde{s}^2~,\quad \dd\tilde{s}^2=4\upsilon \frac{\dd z\dd\bar{z}}{(1+z\bar{z})^2} \equiv e^{2\Omega(z,\bar{z})} \dd z\dd\bar{z}~.
\end{equation}
The action of TLT on a two-sphere topology in Weyl gauge is given by 
\begin{equation}\label{eq:StL}
S_{tL}[\varphi]= \frac{1}{4\pi} \int_{S^2} \dd^2x\sqrt{\tilde{g}} \left( -\tilde{g}^{ij} \partial_i \varphi \partial_j \varphi -  q \tilde{R} \varphi + 4\pi \Lambda e^{2\beta\varphi}  \right)~,
\end{equation}
where $\tilde{R}=2/\upsilon$ is the Ricci scalar of the fiducial metric $\tilde{g}_{ij}$; $\Lambda> 0$ is the cosmological constant. Furthermore we have $q= \beta^{-1}-\beta$. TLT, believed to be a two-dimensional CFT has central charge $c_{tL}= 1-6q^2$. Together with the central charge of the matter theory, coupled in (\ref{eq:StL}) via the identity operator to gravity, and the central charge $c_{\text{gh}}=-26$ of the ghost theory, arising upon gauge fixing, it obeys the condition 
\begin{equation}
c_{tL}+ c_m + c_{\text{gh}} =0~,
\end{equation}
guaranteeing the vanishing of the conformal anomaly. 
The path integral of TLT is
\begin{equation}\label{ZL1}%^{(0)}
\mathcal{Z}_{tL}[\Lambda] =  \frac{1}{\text{vol}_{PSL(2,\mathbb{C})}} \times \int [\mathcal{D}\varphi ] e^{-S_{tL}[\varphi]}~.
\end{equation}
Expanding the Weyl mode into a basis of \text{real} spherical harmonics on the two-sphere
\begin{equation}\label{decomposition}
\varphi(\theta,\phi)=\sum_{l=0}^\infty\sum_{m=-l}^l \varphi_{l,m}Y_{l,m}(\theta,\phi)~, \quad \theta \in [0,\pi)~,\quad \phi \sim \phi +2\pi~.
\end{equation}
we define the measure in (\ref{ZL1})
\begin{equation}
[\mathcal{D}\varphi] = \prod_{l,m} \left(\frac{\Lambda_{\mathrm{uv}}\upsilon}{\pi}\right)^{\frac{1}{2}}{\text{d}  \varphi_{lm}}~.
\end{equation}
such that
% ({\color{magenta} does not work. You need to put $\upsilon\Lambda_{uv}$ upstairs. You sure about the $4\pi$? I think you normlised the spherical harmonics to 1 over $\dd \Omega$})
\begin{equation}
1 = \int [\mathcal{D} \varphi] e^{- \Lambda_{\mathrm{uv}} \int \dd^2 x \sqrt{\tilde{g}} \, \varphi(x)^2}~.
\end{equation}
Here $\Lambda_{\text{uv}}$ is the UV cutoff of the theory.  
\subsection{Fadeev-Popov gauge fixing, round saddle $\&$ small fluctuations}
First and foremost we need to take care of the infinite volume of $PSL(2,\mathbb{C})$ in (\ref{ZL1}). This infinite volume might suggest a vanishing sphere partition function, however the invariance of the Liouville action under \cite{timelike}
\begin{equation}\label{phiT}
\varphi(z,\bar{z}) \to \varphi(f(z),\overline{f(z)}) + \frac{q}{2} \log f'(z) + \frac{q}{2} \log \overline{f'(z)} + q \left(\Omega(f(z),\overline{f(z)} ) - \Omega(z,\bar{z}) \right)~,
\end{equation}
produces another volume of $PSL(2,\mathbb{C})$ upstairs and thus yields an infinity over infinity situation in (\ref{ZL1}). In (\ref{phiT}) $f(z)$ is an element in $PSL(2,\mathbb{C})$ and we defined $\Omega(z,\bar{z})$ in (\ref{metric_start}).
We will fix the volume of $PSL(2,\mathbb{C})$ by using a Fadeev-Popov approach. We follow \cite{Distler:1988jt, timelike} and set the three $l=1$ modes $\delta\varphi_{1,m}$, $m\in \{-1,0,1\}$ to zero. This fixes three of the six parameters of $PSL(2,\mathbb{C})$ and we are left with the finite volume of $SO(3)$. Explicitly we obtain 
\begin{multline}\label{FPexact}
\Delta_{\text{FP}} \equiv \det\frac{\dd \delta\varphi_{1,m}}{\dd \delta\alpha_n}=a_0 q^3 + a_1 q\sum_{m=-2}^2 \varphi^2_{2m} + a_2 \Big(\varphi_{2,0}^3+\frac{3}{2}\varphi_{2,0}(\varphi_{2,1}^2+\varphi_{2,-1}^2)+\frac{3}{2}\sqrt{3}\varphi_{2,2}(\varphi_{2,1}^2-\varphi_{2,-1}^2)\cr
+3\sqrt{3}\varphi_{2,1}\varphi_{2,-1}\varphi_{2,-2}-3\varphi_{2,0}(\varphi_{2,-2}^2+\varphi_{2,2}^2)\Big)~,
\end{multline}
where $m\in \{-1,0,1\}$ and $\alpha_n$ denote directions in $PSL(2,\mathbb{C})/SO(3)$. One can check that (\ref{FPexact}) is indeed $SO(3)$ invariant. Furthermore we find
\begin{equation}\label{a0a1a2}
a_0\equiv -\frac{16}{3\sqrt{3}}\pi^{3/2}~,\quad a_1\equiv \frac{12}{5}\sqrt{3\pi}~,\quad a_2\equiv \frac{12}{5}\sqrt{\frac{3}{5}}~.
\end{equation}
\begin{center}
***
\end{center}
The classical equations of motion of (\ref{eq:StL}) are
\begin{equation}\label{eoms}
-2\tilde{\nabla}^2 \varphi = 8\pi \beta\Lambda e^{2\beta \varphi} -\frac{2}{\upsilon}q~,
\end{equation}
where $-\tilde{\nabla}^2$ is the Laplacian with respect to $\tilde{g}_{ij}$. Equation (\ref{eoms}) admits a constant and real solution given by\footnote{If we allow also complex saddles we have the integer indexed family of constant solutions 
\begin{equation}
\varphi_{c,*}= \varphi_*+ \frac{\pi i}{\beta}n~,\quad n\in \mathbb{Z}~.
\end{equation}
}
\begin{equation}\label{saddle}
\varphi_{*}= \frac{1}{2\beta}\log\left(\frac{q}{4\pi\upsilon\Lambda\beta}\right)~,
\end{equation}
and we obtain the saddle point contribution of (\ref{ZL}) 
\begin{equation}\label{ZLclas}
\mathcal{Z}_{\text{saddle}}[\Lambda] = \left(\frac{q }{4\pi e \Lambda \upsilon \beta } \right)^{\frac{q}{\beta}} \approx  \left(\frac{1}{\Lambda \upsilon \beta^2}\right)^{\frac{1}{\beta^2}}~.
\end{equation}
We now add a small fluctuation $\delta \varphi$ to $\varphi_*$
\begin{equation}\label{fluctuations}
\varphi\rightarrow \varphi_*+\delta\varphi~.
\end{equation}
Using the expansion (\ref{decomposition}) the Laplacian $\tilde{\nabla}^2$ for $\delta\varphi$ obeys the eigenvalue equation 
\begin{equation}
\tilde{\nabla}^2\delta\varphi=- \frac{1}{\upsilon}l(l+1)\delta\varphi~,\quad l \geq 0~.
\end{equation}
Expanding (\ref{eq:StL}) around (\ref{fluctuations}) we observe that for $l\geq 2$ we are dealing with an infinite number of unsuppressed Gaussian terms. To cure this we will follow \cite{Gibbons:1978ac}  and rotate $\delta\varphi \rightarrow \pm i \delta\varphi$. The resulting Jacobian is an ultralocal contribution \cite{Polchinski:1988ua} and can be absorbed in the measure. 
We thus have
\begin{equation}
\mathcal{Z}_{tL}[\Lambda]= \mathcal{Z}_{\text{saddle}}[\Lambda]\times \mathcal{Z}_{\text{pert}}[\beta]~,
\end{equation}
where $\mathcal{Z}_{\text{saddle}}$ we defined in (\ref{ZLclas}) and the perturbative part is given by 
\begin{multline}\label{Zpert2}
\mathcal{Z}_{\text{pert}}[\beta]= \frac{1}{\text{vol}_{SO(3)}}\times  \int\,[\mathcal{D} \delta\varphi ] \times  \Delta_{\text{FP}}[\delta\varphi]\times \prod_{m=\{-1,0,1\}} \delta(\delta\varphi_{1m}) \times e^{- \frac{1}{4\pi} \int_{S^2} \dd^2x \sqrt{\tilde{g}} \left(\tilde{g}^{ij} \partial_i \delta \varphi \partial_j \delta \varphi - \frac{2}{\upsilon}q\beta \delta \varphi^2  \right)}\cr
\times e^{-\frac{1}{4\pi}\frac{q}{\beta}\int \dd \Omega\left(- i\frac{4}{3}\beta^3 \delta\varphi(\Omega)^3+ \frac{2}{3}\beta^4 \delta\varphi(\Omega)^4+ i\frac{4}{15}\beta^5\delta\varphi(\Omega)^5- \frac{4}{45}\beta^6\delta\varphi(\Omega)^6+ \ldots\right)}~.
\end{multline}
%The quadratic action is
%\begin{equation}\label{Spert2}
%S_{\text{pert}}^{(2)}[\delta \varphi ]  = \frac{1}{4\pi} \int \dd^2x \sqrt{\tilde{g}} \left(\tilde{g}^{ij} \partial_i \delta \varphi \partial_j \delta \varphi - \frac{2}{\upsilon}q\beta \delta \varphi^2  \right)~.
%\end{equation}
\subsection{One-loop contribution}
To quadratic order in the fields and in the semiclassical small $\beta$ limit the path integral (\ref{Zpert2}) including the Fadeev-Popov determinant is given by
\begin{equation}\label{Zpert}
\mathcal{Z}^{(2)}_{\text{pert}}[\beta]=a_0q^3 \int \prod_{l,m}\left(\frac{\Lambda_{\text{uv}}\upsilon}{\pi}\right)^{\frac{1}{2}} \dd\delta\varphi_{l,m}\times \prod_{m=\{-1,0,1\}} \delta(\delta\varphi_{1m})\,e^{\frac{a_1}{a_0q^2}\sum_{m=-2}^{2}\delta\varphi_{2,m}^2}e^{- \frac{1}{4\pi}\sum_{l,m}\left(l(l+1)-2\beta q\right)\delta\varphi_{l,m}^2}~.
\end{equation}
We highlight that we also Wick-rotated the $l=2$ mode in the Fadeev-Popov determinant (\ref{FPexact}) and consequently we have in the above expression
\begin{equation}\label{a1a0}
\frac{a_1}{a_0}= + \frac{27}{20\pi}~.
\end{equation}
Whereas the Wick rotation $\delta\varphi \rightarrow \pm i \delta\varphi$ cured the unsuppressed Gaussians for $l \geq 2$, from (\ref{Zpert}) we infer that it created a Gaussian unsuppressed $l=0$ mode. We cure this by Wick rotating a single mode back $\delta \varphi_{00} \rightarrow \pm i\delta\varphi_{00}$  \cite{Gibbons:1978ac,Polchinski:1988ua} . Keeping track of the resulting Jacobian we arrive at
\begin{equation}\label{Zpert22}
\mathcal{Z}^{(2)}_{\text{pert}}[\beta] \equiv \pm i a_0 q^3\left(\frac{2\pi\upsilon\Lambda_{\mathrm{uv}}}{\beta q}\right)^{\frac{1}{2}}  \left({\frac{\upsilon\Lambda_{\mathrm{uv}}}{\pi}}\right)^{\frac{3}{2}}\left(\frac{4\pi\upsilon\Lambda_{\mathrm{uv}}}{6-2\beta q -4\pi\frac{a_1}{a_0q^2}}\right)^{\frac{5}{2}}\prod^\infty_{l=3}\left(\frac{4\pi\upsilon\Lambda_{\mathrm{uv}}}{l(l+1)-2\beta q}\right)^{l+\frac{1}{2}}~.
% \int [\mathcal{D}\varphi] e^{-S^{(2)}[\delta \varphi ]}
\end{equation}
In the above expression we have been treating the $l=0,1,2$ and $l\geq 3$ modes separately. In particular for $l\geq 3$ we encounter an infinite product which we can evaluate using for example a heat kernel regularisation scheme. We obtain
\begin{multline}\label{heatkernel}
-\frac{1}{2} \sum_{l=3}^\infty ({2l+1} ) \log \left( \frac{l(l+1) -2\beta q}{4\pi\Lambda_{\mathrm{uv}} \upsilon} \right) = -\frac{107+12\nu^2}{12}\log\left(\frac{2e^{-\gamma_E}}{\varepsilon}\right)+\frac{2}{\varepsilon^2}+\nu^2\cr
+\left(\frac{1}{2}-\Delta_+\right)\zeta'(0,\Delta_+)+ \left(\frac{1}{2}-\Delta_-\right)\zeta'(0,\Delta_-)+\zeta'(-1,\Delta_+)+\zeta'(-1,\Delta_-)~\cr
+\frac{3}{2}\log \beta^2+\frac{5}{2}\log(2+\beta^2)+\frac{1}{2}\log (-1+\beta^2) + \frac{9}{2}\log 2~.
%\\ \frac{2}{\varepsilon^2} + \left(\frac{1}{12}-\nu^2\right) \log \frac{2e^{-\gamma}}{{\varepsilon}} +  \sum_{\Delta = \frac{1}{2} \pm i \nu} \left( \zeta'(-1,\Delta) - (\Delta-1/2)\zeta'(0,\Delta) \right) + \nu^2~,
\end{multline}
where $\nu \equiv \sqrt{-2\beta q-1/4}$, $\Delta_{\pm}= 1/2\pm i\nu$, and $\zeta(a,z)$ denotes the Hurwitz $\zeta$-function. Furthermore we have $\varepsilon= e^{-\gamma_E}/\sqrt{\pi \upsilon\Lambda_{\text{uv}}}$. Applying to (\ref{heatkernel}) the relations \cite{Adamchik}\footnote{These identities are to be understood as yielding a real valued analytic expression at small $\beta$.} 
\begin{equation}
\zeta'(0,z)= \text{log}\Gamma(z)-\frac{1}{2}\log(2\pi)~,\quad \zeta'(-1,z)= \zeta'(-1) -\log G(z+1)+z \, \text{log}\Gamma(z)~,
\end{equation} 
we find
\begin{align}\allowdisplaybreaks
&+\left(\frac{1}{2}-\Delta_+\right)\zeta'(0,\Delta_+)+ \left(\frac{1}{2}-\Delta_-\right)\zeta'(0,\Delta_-)+\zeta'(-1,\Delta_+)+\zeta'(-1,\Delta_-)~\cr
&+\frac{1}{2}\log(-1+\beta^2)+\frac{3}{2}\log \beta^2+\frac{5}{2}\log(2+\beta^2)+ \frac{9}{2}\log 2\cr
&\sim \frac{1}{6}+\frac{1}{2}\log(6)+\log(96)-2\log A+\left(\frac{13}{12}-2\gamma\right)\beta^2+\frac{61}{432}\beta^4+\mathcal{O}\left(\beta^6\right)~.
\end{align}
For details to the calculation of the Fadeev-Popov determinant and the heat-kernel analysis we refer to \cite{timelike}.
\begin{center}
***
\end{center}
To conclude this section we discuss the effect of the Fadeev-Popov determinant (\ref{FPexact}). Expanding the contribution of the Fadeev-Popov determinant for small $\beta$ and stripping off the factor $a_0 q^3$ we obtain\footnote{Note that we could have also calculated 
\begin{equation}
\int \prod_{m=-2}^{2}\dd \delta\varphi_{2,m}\, \Delta_{\text{FP}}\, e^{-\frac{1}{4\pi}(6-2\beta q)\delta\varphi_{2,m}^2+\ldots}~,
\end{equation}
explicitly. We decided to calculate the pieces order by order to have a better idea about the propagator. 
} 
\begin{align}\label{FP_contribution}
\mathcal{Z}_{\text{pert}}^{\text{FP}}[\beta]&=a_0 q^3 \int \prod_{m=-2}^2 \dd \varphi_{2,m}\,e^{-\frac{1}{4\pi}(6-2q\beta)}\times e^{\log \frac{\Delta_{\text{FP}}}{a_0 q^3}}\cr
&= a_0 q^3\int \prod_{m=-2}^2 \dd \varphi_{2,m}\, e^{-\frac{1}{4\pi}\left(6-2\beta q- 4\pi \frac{a_1}{a_0q^2}\right)\delta\varphi_{2,m}^2}\, e^{\frac{a_2}{a_0}\Phi_3\beta^3-\frac{1}{2}\left(\frac{a_1}{a_0}\right)^2\Phi_2^2 \beta^4+ \ldots}~,
%&=  \int[\mathcal{D}' \delta\varphi]\, e^{-S^{(2)}_{\text{pert}}[\delta\varphi]}\, \left(1- \frac{1}{2}\left(\frac{a_1}{a_0}\right)^2\left(\sum_{m=-2}^2 \varphi^2_{2m}\right)^2\beta^4\right)
\end{align}
where for notational convenience we defined 
\begin{align}
\Phi_2&\equiv \sum_{m=-2}^2 \delta\varphi_{2,m}^2~,\quad\Phi_3\equiv \delta\varphi_{2,0}^3+\frac{3}{2}\delta\varphi_{2,0}(\delta\varphi_{2,1}^2+\delta\varphi_{2,-1}^2)+\frac{3}{2}\sqrt{3}\delta\varphi_{2,2}(\delta\varphi_{2,1}^2-\delta\varphi_{2,-1}^2)
\cr &+3\sqrt{3}\delta\varphi_{2,1}\delta\varphi_{2,-1}\delta\varphi_{2,-2}-3\delta\varphi_{2,0}(\delta\varphi_{2,-2}^2+\delta\varphi_{2,2}^2)~.
\end{align}
Expanding the exponential, we obtain to order $\mathcal{O}(\beta^4)$ 
\begin{equation}\label{FP_diagrams}
\frac{1}{\mathcal{Z}_{\text{pert}}^{(2)}[\beta]} \times \mathcal{Z}_{\text{pert}}^{\text{FP}}[\beta]
%= 1 -\frac{70\pi^2}{A_2^2}\left(\frac{a_1}{a_0}\right)^2\beta^4+ \left(\frac{840\pi^3}{A_2^3}\left(\frac{a_1}{a_0}\right)^3- \frac{240\pi^2}{A_2^2}\left(\frac{a_1}{a_0}\right)^2- \frac{420\pi^3}{A_2^3}\left(\frac{a_2}{a_0}\right)^2\right)\beta^6+ \ldots\cr
= 1 -\frac{35\pi^2}{8}\left(\frac{a_1}{a_0}\right)^2\beta^4+ \ldots~.
\end{equation}
In particular we note the first appearance of $a_2$ will be at order $\mathcal{O}(\beta^6)$ and that we absorbed the phases of the Wick rotation of the FP determinant into the coefficients $a_i$ (\ref{a0a1a2}).
%\begin{equation}
%\frac{1}{\mathcal{Z}_{\text{pert}}^2}\times \mathcal{Z}_{\text{pert}}^{\text{FP}}[\beta]= 1- 
%\end{equation}
\section{Diagramatics}\label{sec:3}
In this section we calculate the higher-loop corrections to (\ref{Zpert2}). We start by explaining the propagator on the two-sphere, review the two-loop calculations of \cite{timelike} and then delve into the three loop-contributions. 
\subsection{Propagators $\&$ spherical harmonics}
First we note the propagator. For $\Omega, \Omega'$ two points on the round two-sphere we have
\begin{equation}\label{propagator}
G(\Omega;\Omega') \equiv  \frac{1}{\mathcal{Z}_{\mathrm{pert}}^{(2)}[\beta]} \int [\mathcal{D}'\delta\varphi] e^{-S_{\mathrm{pert}}^{(2)}[\delta\varphi]} \delta\varphi(\Omega) \delta\varphi(\Omega') = 2\pi \sum_{l\neq 1 ,m\in[-l,l]} \frac{{Y}_{l m}(\Omega) {Y}_{l m}(\Omega')}{A_l} ~,
%\langle \phi(\Omega)\phi(\Omega') \rangle = 
\end{equation}
where we defined for $l\neq 1$, $A_l\equiv (l(l+1)-2+2\beta^2- 4\pi a_1 /(a_0 q^2) \delta_{l,2})$, $\Omega$ is a point on the round two-sphere and $Y_{l,m}(\Omega)$ denote the real spherical harmonics. Our conventions we explain in appendix \ref{Yapp}.
In particular at coincidence where $\Omega=\Omega'$ we have 
\begin{equation}\label{divloop}
\int_{S^2}\dd\Omega G(\Omega;\Omega) = {2\pi} \sum_{l\neq 1}^\infty \frac{2l+1}{A_l} = 4\pi G(\Omega_0;\Omega_0)~.
\end{equation}
The last equality follows from the fact that $G(\Omega;\Omega)$ is $\Omega$ independent. The above sum diverges logarithmically, as expected for coincident fields in two dimensions. (\ref{divloop}) holds true also if we remove the $l=1$ modes since for each $l$ the spherical harmonics form an irreducible representation of $SO(3)$. Finally for the $l=2$ modes we also need to take into consideration the effects arising from the Fadeev-Popov determinant (\ref{Zpert}), thus shifting the ``mass'' in $A_l$. We now examine the Feynman diagrams in (\ref{Zpert2}). We first review the two-loop calculation in \cite{timelike} and then delve into the three-loop contributions. Care must be taken since we need to remove the $l=1$ modes and adjust the propagator for the $l=2$ modes.
\subsection{Two-loop contributions}
We now discuss the path integral (\ref{Zpert2})
\begin{align}
\mathcal{Z}_{\text{pert}}[\beta]&= \int\,[\mathcal{D} \delta\varphi ] \times  \Delta_{\text{FP}}[\delta\varphi]\times \prod_{m=\{-1,0,1\}} \delta(\delta\varphi_{1m}) \cr
&\times e^{-S^{(2)}_{\text{pert}}[\delta\varphi]}e^{-\frac{1}{4\pi}\frac{q}{\beta}\int_{S^2} \dd \Omega\left(- i\frac{4}{3}\beta^3 \delta\varphi(\Omega)^3+ \frac{2}{3}\beta^4 \delta\varphi(\Omega)^4+ i\frac{4}{15}\beta^5\delta\varphi(\Omega)^5- \frac{4}{45}\beta^6\delta\varphi(\Omega)^6+ \ldots\right)}
\end{align}
in a small $\beta$ expansion. We remind the reader of the relation $q= \beta^{-1}-\beta$. The leading two-loop contribution has already been calculated in \cite{timelike}. We summarise the main results. To two-loop order we obtain three different types of diagrams which we denote as double tadpoles, cactus diagrams and melonic type diagrams (see fig. \ref{fig:melons_tadpoles_vector})
\begin{align}\label{melons}
\bigcirc\!\!-\!\!\bigcirc\equiv &-\frac{1}{A_0}\sum_{\boldsymbol{l}\neq \bold{1}}\frac{(2l_1+1)(2l_2+1)}{A_{l_1}A_{l_2}}~,\quad
\bigcirc\!\bigcirc\equiv  -\frac{1}{2}\sum_{\boldsymbol{l}\neq \bold{1}}\frac{(2l_1+1)(2l_2+1)}{A_{l_1}A_{l_2}}~,\cr
\mathlarger{\mathlarger{\ominus}}\equiv &-\frac{2}{3}\sum_{\boldsymbol{l}\neq \bold{1}}\frac{(2l_1+1)(2l_2+1)(2l_3+1)}{A_{l_1}A_{l_2}A_{l_3}}\begin{pmatrix} l_1 & l_2 & l_3 \\ 0 & 0 & 0\end{pmatrix}^2~.
\end{align}
\begin{figure}[H]
\begin{center}
\begin{tikzpicture}[scale=.4]

\draw[black] (-1,0) --(-.5,0);
\draw (0.5,0) circle (1cm);
\draw (-2.,0) circle (1cm);

\draw (6.5,0) circle (1cm);
\draw (8.5,0) circle (1cm);

\draw (14.5,0) circle (1cm);
\draw[black] (13.5,0) --(15.5,0);

\end{tikzpicture}
\end{center}
\caption{Double-tadpoles, cactus and melonic type diagrams.}
\label{fig:melons_tadpoles_vector}
\end{figure}
\noindent
We use bold symbols to combine summation indices, e.g. in the melonic sum $\mathlarger{\mathlarger{\ominus}}$ the boldsymbol $\boldsymbol{l}= \{l_1,l_2,l_3\}$ and so on. 
Since at order $\mathcal{O}(1)$, $A_0=-2$ we observe that the UV divergent double-tadpole and cactus diagrams cancel each other at order $\mathcal{O}(\beta^2)$. The remaining melonic type diagrams give a UV finite contribution at order $\mathcal{O}(\beta^2)$. In summary the diagramatics so far leads to 
\begin{equation}
\frac{1}{\mathcal{Z}_{\mathrm{pert}}^{(2)}[\beta]} \times \mathcal{Z}_{\text{pert}}[\beta]= 1+ \beta^2 \mathlarger{\mathlarger{\ominus}}+ \ldots
\end{equation}
Due to the $\beta$ dependency of 
\begin{equation}\label{eq:Al}
A_l = l(l+1)-2+2\beta^2 -\frac{4\pi}{q^2}\frac{a_1}{a_0}~,\quad q=\beta^{-1}-\beta~
\end{equation}
however cactus and double-tadpoles do not cancel anymore at order $\mathcal{O}(\beta^4)$. Furthermore for the $l=2$ there is also the effect of the Fadeev-Popov determinant. 
Before delving into three-loop contributions which add diagrams of order $\mathcal{O}(\beta^4)$ we therefore expand $A_l$ in (\ref{melons}):
\begin{multline}\label{eq:twoloop_hO}
\beta^2\left(\bigcirc\!\!-\!\!\bigcirc+\bigcirc\!\!\bigcirc+ \mathlarger{\mathlarger{\ominus}}\right)
%&-\frac{\beta^2}{A_0}\sum_{\boldsymbol{l}\neq 1}\frac{(2l_1+1)(2l_2+1)}{A_{l_1}A_{l_2}}-\frac{\beta^2}{2}\sum_{\boldsymbol{l}\neq 1}\frac{(2l_1+1)(2l_2+1)}{A_{l_1}A_{l_2}}-\frac{2}{3}\beta^2\sum_{\bold{l}\neq 1}\frac{(2l_1+1)(2l_2+1)(2l_3+1)}{A_{l_1}A_{l_2}A_{l_3}}\begin{pmatrix} l_1 & l_2 & l_3 \\ 0 & 0 & 0\end{pmatrix}^2\cr
%&=\frac{\beta^2}{2}\sum_{\boldsymbol{l}\neq 1}\frac{(2l_1+1)(2l_2+1)}{A_{l_1}A_{l_2}} - \frac{\beta^2}{2}\sum_{\boldsymbol{l}\neq 1}\frac{(2l_1+1)(2l_2+1)}{A_{l_1}A_{l_2}}+\frac{\beta^4}{2}\sum_{\boldsymbol{l}\neq 1}\frac{(2l_1+1)(2l_2+1)}{A_{l_1}A_{l_2}}\cr
%&-\frac{2}{3}\beta^2\sum_{\bold{l}\neq 1}\frac{(2l_1+1)(2l_2+1)(2l_3+1)}{A_{l_1}A_{l_2}A_{l_3}}\begin{pmatrix} l_1 & l_2 & l_3 \\ 0 & 0 & 0\end{pmatrix}^2+4\beta^4\sum_{\bold{l}\neq 1}\frac{(2l_1+1)(2l_2+1)(2l_3+1)}{A_{l_1}^2A_{l_2}A_{l_3}}\begin{pmatrix} l_1 & l_2 & l_3 \\ 0 & 0 & 0\end{pmatrix}^2\cr
%&+ \frac{40\pi}{A_2^2}\frac{a_1}{a_0}\beta^4 \sum_{\bold{l}\neq 1,2}\frac{(2l_2+1)(2l_3+1)}{A_{l_2}A_{l_3}}\begin{pmatrix} 2 & l_2 & l_3 \\ 0 & 0 & 0\end{pmatrix}^2+ \frac{200\pi}{A_2^3}\frac{a_1}{a_0}\beta^4 \sum_{\bold{l}\neq 1,2}\frac{(2l_3+1)}{A_{l_3}}\begin{pmatrix} 2 & 2 & l_3 \\ 0 & 0 & 0\end{pmatrix}^2\cr
%&+ \frac{8\pi}{3A_{2}^4}5^3\frac{a_1}{a_0}\beta^4 \begin{pmatrix} 2 & 2 & 2 \\ 0 & 0 & 0\end{pmatrix}^2\cr
=\beta^2\mathlarger{\mathlarger{\ominus}}+ \frac{\beta^4}{2}\sum_{\boldsymbol{l}\neq \bold{1}}\frac{(2l_1+1)(2l_2+1)}{A_{l_1}A_{l_2}}+4\beta^4\sum_{\boldsymbol{l}\neq \bold{1}}\frac{(2l_1+1)(2l_2+1)(2l_3+1)}{A_{l_1}^2A_{l_2}A_{l_3}}\begin{pmatrix} l_1 & l_2 & l_3 \\ 0 & 0 & 0\end{pmatrix}^2\cr
+ \frac{2}{3}\times 3\times 5\times \frac{4\pi}{A_2^2}\frac{a_1}{a_0}\beta^4 \sum_{\boldsymbol{l}\neq 1,2}\frac{(2l_2+1)(2l_3+1)}{A_{l_2}A_{l_3}}\begin{pmatrix} 2 & l_2 & l_3 \\ 0 & 0 & 0\end{pmatrix}^2\cr
+\frac{2}{3}\times 6\times 5^2\times \frac{4\pi}{A_2^3}\frac{a_1}{a_0}\beta^4 \sum_{l_3\neq 1,2}\frac{(2l_3+1)}{A_{l_3}}\begin{pmatrix} 2 & 2 & l_3 \\ 0 & 0 & 0\end{pmatrix}^2
+\frac{2}{3}\times 3\times 5^3\times \frac{4\pi}{A_{2}^4}\frac{a_1}{a_0}\beta^4 \begin{pmatrix} 2 & 2 & 2 \\ 0 & 0 & 0\end{pmatrix}^2~.
\end{multline}
Since the 3j symbol obeys the triangle condition we can evaluate the last two contributions explicitly.
%\begin{multline}\label{eq:twoLoop_beta4}
%\beta^2\left(\bigcirc\!\!-\!\!\bigcirc+\bigcirc\!\!\bigcirc+ \mathlarger{\mathlarger{\ominus}}\right)= \beta^2\mathlarger{\mathlarger{\ominus}}+ \frac{\beta^4}{2}\sum_{\boldsymbol{l}\neq 1}\frac{(2l_1+1)(2l_2+1)}{A_{l_1}A_{l_2}}\cr
%+4\beta^4\sum_{\boldsymbol{l}\neq 1}\frac{(2l_1+1)(2l_2+1)(2l_3+1)}{A_{l_1}^2A_{l_2}A_{l_3}}\begin{pmatrix} l_1 & l_2 & l_3 \\ 0 & 0 & 0\end{pmatrix}^2{\color{magenta}+\frac{5477\pi}{50400}\frac{a_1}{a_0}\beta^4}- \frac{25\pi}{112}\frac{a_1}{a_0}\beta^4+\frac{25\pi}{448}\frac{a_1}{a_0}\beta^4~.
%\end{multline}
Clearly then only the second contribution in (\ref{eq:twoloop_hO}) is logarithmically divergent in the UV whereas the other diagrams yield finite contribution. 

We now start our journey calculating the three-loop contributions, thereby finding a contribution exactly and fully cancelling the logarithmically divergent contribution in (\ref{eq:twoloop_hO}).

\subsection{Three-loop contributions}\label{sec:33}
To three loop and at order $\mathcal{O}(\beta^4)$ we obtain from (\ref{Zpert2}) the following seven contributions
\begin{multline}\label{3loop_diagrams1}
{\color{black}\frac{1}{9\pi^2}\bigg\langle\int_{S^2} \dd\Omega \dd\Omega' \varphi(\Omega)^3\varphi(\Omega')^3}\bigg\rangle+{\color{black}\frac{1}{6\pi}\bigg\langle\int_{S^2} \dd\Omega \varphi(\Omega)^4}\bigg\rangle+{\color{black} \frac{1}{45\pi^2}\bigg\langle\int_{S^2} \dd\Omega \dd\Omega' \varphi(\Omega)^3\varphi(\Omega')^5}\bigg\rangle\cr
+ {\color{black}\frac{1}{45\pi}\bigg\langle\int_{S^2} \dd\Omega \varphi(\Omega)^6}\bigg\rangle
+{\color{black}\frac{1}{72\pi^2} \bigg\langle\int_{S^2} \dd\Omega\dd\Omega'\varphi(\Omega)^4\varphi(\Omega')^4}\bigg\rangle+ {\color{black}\frac{1}{108\pi^3}\bigg\langle\int_{S^2} \dd\Omega \dd\Omega'\dd\Omega'' \varphi(\Omega)^3\varphi(\Omega')^3\varphi(\Omega'')^4}\bigg\rangle\cr
+{\color{black} \frac{1}{1944\pi^4}\bigg\langle\int_{S^2} \dd\Omega \dd\Omega'\dd\Omega''\dd{\Omega'''} \varphi(\Omega)^3\varphi(\Omega')^3\varphi(\Omega'')^3\varphi(\Omega''')^3}\bigg\rangle~.
\end{multline}
Here $\langle \cdot \rangle $ denotes the expectation value with respect to the Gaussian integral (\ref{Zpert}).
We explain the diagrams individually.

First we highlight that two-loop diagrams appear at order $\mathcal{O}(\beta^4)$ not only through the $\beta$ dependency in $A_l$ (\ref{eq:Al}) (see (\ref{eq:twoloop_hO})) but also due to the fact that the couplings in (\ref{Zpert2}) are of the form $q\beta^{a}, a\geq 2$, which itself has a small $\beta$ expansion.
The first two diagrams in (\ref{3loop_diagrams1}) are thus well known, adding melons, double-tadpoles and vector diagrams:
\begin{multline}\label{eq:phi32}
\frac{1}{9\pi^2}\bigg\langle\int_{S^2} \dd\Omega \dd\Omega' \varphi(\Omega)^3\varphi(\Omega')^3\bigg\rangle=\frac{1}{9\pi^2} {\color{black}c_1}\times \int_{S^2} \dd\Omega\dd\Omega' G(\Omega,\Omega')^3+ \frac{1}{9\pi^2}{\color{black}c_2}\times G(\Omega_0,\Omega_0)^2 \int_{S^2} \dd\Omega\dd\Omega' G(\Omega,\Omega')~\cr
= \frac{4}{3}\sum_{\boldsymbol{l}\neq \bold{1}}\frac{(2l_1+1)(2l_2+1)(2l_3+1)}{A_{l_1}A_{l_2}A_{l_3}}\begin{pmatrix} l_1 & l_2 & l_3 \\ 0 & 0 & 0\end{pmatrix}^2
+ \frac{2}{A_0}\sum_{\boldsymbol{l}\neq \bold{1}}\frac{(2l_1+1)(2l_2+1)}{A_{l_1}A_{l_2}}~,
\end{multline}
where $c_1=6$ and $c_2=9$ are two combinatorial factors. So we find melons and double-tadpoles also at three-loop order. 
The second contribution in (\ref{3loop_diagrams1})  are cactus diagrams
\begin{equation}\label{eq:phi4}
\frac{1}{6\pi}\bigg\langle\int_{S^2} \dd\Omega \varphi(\Omega)^4\bigg\rangle=\frac{1}{6\pi}c_1\times \int_{S^2} \dd\Omega G(\Omega,\Omega)^2= \frac{1}{2}\sum_{\boldsymbol{l}\neq \bold{1}}\frac{(2l_1+1)(2l_2+1)}{A_{l_1}A_{l_2}}~.
\end{equation}
Now we delve into the world of three-loop contributions. We have
\begin{align}\label{eq:phi3phi5}
&\frac{1}{45\pi^2}\bigg\langle\int_{S^2} \dd \Omega \dd\Omega' \varphi(\Omega)^3\varphi(\Omega')^5\bigg\rangle = \frac{1}{45\pi^2}{\color{black}c_1}\times G(\Omega_0,\Omega_0)\int_{S^2} \dd\Omega \dd\Omega' G(\Omega,\Omega')^3\cr
&+ \frac{1}{45\pi^2}{\color{black}c_2} \times G(\Omega_0,\Omega_0)^3\int_{S^2} \dd\Omega \dd\Omega' G(\Omega,\Omega')=\frac{4}{3}\sum_{\boldsymbol{l}\neq \bold{1}}\frac{(2l_1+1)(2l_2+1)(2l_3+1)(2l_4+1)}{A_{l_1}A_{l_2}A_{l_3}A_{l_4}}\begin{pmatrix} l_1 & l_2 & l_3 \\ 0 & 0 & 0\end{pmatrix}^2\cr
&+ \frac{1}{A_0} \sum_{\boldsymbol{l}\neq \bold{1}} \frac{(2l_1+1)(2l_2+1)(2l_3+1)}{A_{l_1}A_{l_2}A_{l_3}}~.
\end{align}
The combinatorial factors are $c_1=60$ and $c_2=45$; The sextic diagram evaluates to 
\begin{equation}\label{eq:phi6}
\frac{1}{45\pi} \bigg\langle\int_{S^2} \dd\Omega \varphi(\Omega)^6\bigg\rangle = \frac{1}{6} \sum_{\boldsymbol{l}\neq \bold{1}}\frac{(2l_1+1)(2l_2+1)(2l_3+1)}{A_{l_1}A_{l_2}A_{l_3}}~.
\end{equation}
Now we are entering more subtle ground. We find
\begin{multline}\label{eq:phi42}
\frac{1}{72\pi^2} \bigg\langle\int_{S^2} \dd\Omega\dd\Omega'\varphi(\Omega)^4\varphi(\Omega')^4\bigg\rangle=  \sum_{\boldsymbol{l}\neq \bold{1}}\frac{(2l_1+1)(2l_2+1)(2l_3+1)}{A_{l_1}A_{l_2}A_{l_3}^2}+\frac{1}{8} \sum_{\boldsymbol{l}\neq \bold{1}}\frac{(2l_1+1)(2l_2+1)(2l_3+1)(2l_4+1)}{A_{l_1}A_{l_2}A_{l_3}A_{l_4}}\cr
+ \frac{1}{3} \sum_{\boldsymbol{l}\neq \bold{1}}\sum_{l_5\geq 0}\frac{(2l_1+1)(2l_2+1)(2l_3+1)(2l_4+1)(2l_5+1)}{A_{l_1}A_{l_2}A_{l_3}A_{l_4}}\begin{pmatrix} l_1 & l_2 & l_5\\ 0 & 0 & 0\end{pmatrix}^2\begin{pmatrix} l_3 & l_4 & l_5\\ 0 & 0 & 0\end{pmatrix}^2~,
\end{multline}
where we highlight that in the last sum the sum over $l_5$ runs over all positive integers subject only to the triangle condition of the 3j symbols (see appendix \ref{Yapp}). 
Finally we are dealing with the last two terms in (\ref{3loop_diagrams1}). The evaluation of these diagrams is a bit more cumbersome and we refer to appendix \ref{appendix_Loops} for the results. Graphically the above diagrams correspond to 
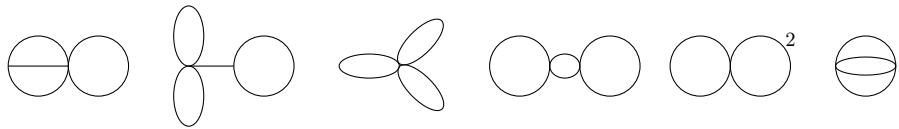
\begin{figure}[H]
\begin{center}
\begin{tikzpicture}[scale=.4]

\draw (-2.5,1) circle (1cm);
\draw (-.5,1) circle (1cm);
\draw[black] (-3.5,1) --(-1.5,1);

\draw (2.5,2) ellipse (.5cm and 1cm);
\draw (2.5,0) ellipse (.5cm and 1cm);
\draw[black] (2.5,1) --(4,1);
\draw (5,1) circle (1cm);

\draw (8.5,1) ellipse (1cm and .4cm);
\draw[rotate around={-45:(10.2,1.8)}] (10.2,1.8) ellipse (.4cm and 1cm);
\draw[rotate around={45:(10.215,0.28)}] (10.215,0.28) ellipse (.4cm and 1cm);

%\draw (14,1) circle (1cm);
%\draw[black] (15,1) --(16,1.8);
%\draw[black] (15,1) --(16,.2);
%\draw[black] (16,1.8) --(16,.2);
%\draw (16,1.8) to[out=20,in=-10] (16,.2);
%\draw (-4,0) circle (.5cm);
%\draw (-2.5,0) circle (1cm);

\draw (13.5,1) circle (1cm);
\draw[] (15,1) ellipse (.5cm and .4cm);
\draw (16.5,1) circle (1cm);

\draw (19.5,1) circle (1cm);
\draw (21.5,1) circle (1cm);
\node[scale=.7] at (22.5,1.9) {2};
%\draw (19.5,1) ellipse (1cm and .2cm);
%\draw[] (20.5,2) ellipse (.2cm and 1cm);
%\draw (20.5,0) ellipse (.2cm and 1cm);
%\draw (21.5,1) ellipse (1cm and .2cm);

\draw (25,1) circle (1cm);
\draw (25,1) ellipse (1cm and .3cm);

\end{tikzpicture}

\end{center}
\caption{Diagrams in order of appearance starting at (\ref{eq:phi3phi5}).}
\label{fig:diagramsValpha}
\end{figure}

\section{Sphere partition function}\label{sec:spherePI}
We now combine the 26 diagrams (\ref{eq:phi32}--\ref{eq:phi42}) and (\ref{eq:phi32phi4}--\ref{eq:phi34}) with the $\mathcal{O}(\beta^4)$ contribution of the two-loop diagrams (\ref{eq:twoloop_hO}). 
\subsection{Cancellations}
\begin{figure}[H]
\begin{center}
\begin{tikzpicture}[scale=.3]

\draw (-7.5,1) circle (1cm);
\draw (-5.5,1) circle (1cm);

\node[scale=.5] at (-4,1) {+};

\draw (-2.5,1) circle (1cm);
\draw[black] (-1.5,1) --(-1,1);
\draw (0,1) circle (1cm);

\node[scale=.5] at (1.8,1) {+};

\draw (3,2) ellipse (.5cm and 1cm);
\draw (3,0) ellipse (.5cm and 1cm);
\draw[black] (3,1) --(4,1);
\draw (5.,1) circle (1cm);

\node[scale=.5] at (6.8,1) {+};

\draw (8.5,1) ellipse (1cm and .4cm);
\draw[rotate around={-45:(10.2,1.8)}] (10.2,1.8) ellipse (.4cm and 1cm);
\draw[rotate around={45:(10.215,0.28)}] (10.215,0.28) ellipse (.4cm and 1cm);

%\draw (14,1) circle (1cm);
%\draw[black] (15,1) --(16,1.8);
%\draw[black] (15,1) --(16,.2);
%\draw[black] (16,1.8) --(16,.2);
%\draw (16,1.8) to[out=20,in=-10] (16,.2);
%\draw (-4,0) circle (.5cm);
%\draw (-2.5,0) circle (1cm);

\node[scale=.5] at (11.8,1) {+};

\draw (13.5,1) circle (1cm);
\draw (16,2.2) circle (1cm);
\draw (16,-.2) circle (1cm);
\draw[black] (14.5,1) --(15.2,1.65);
\draw[black] (14.5,1) --(15.2,.35);

%\draw (13.5,1) circle (1cm);
%\draw[] (15,1) ellipse (.5cm and .4cm);
%\draw (16.5,1) circle (1cm);

\node[scale=.5] at (18,1) {+};

\draw (19.5,1) circle (1cm);
\draw (22.5,2.2) circle (1cm);
\draw (22.5,-.2) circle (1cm);
\draw[black] (20.5,1) --(21.,1);
\draw[black] (21.,1) --(21.7,1.65);
\draw[black] (21.,1) --(21.7,.35);

\node[scale=.5] at (23.8,1) {+};

\draw (25.5,1) circle (1cm);
\draw (27.5,1) circle (1cm);
\draw (30,1) circle (1cm);
\draw[black] (28.5,1) --(29,1);

\node[scale=.5] at (31.6,1) {+};

\draw (33,1) circle (1cm);
\draw[] (34.5,1) ellipse (.5cm and .4cm);
\draw (36,1) circle (1cm);

%\draw (32.5,1) circle (1cm);
%\draw (36,2.2) circle (1cm);
%\draw (36,-.2) circle (1cm);
%\draw[black] (33.5,1) --(34.5,1);
%\draw[black] (34.5,1) --(35.2,1.65);
%\draw[black] (34.5,1) --(35.2,.35);

\node[scale=.5] at (37.7,1) {+};

\draw (39.5,1) circle (1cm);
\draw (42,1) circle (1cm);
\draw (44.5,1) circle (1cm);
\draw[black] (40.5,1) --(41,1);
\draw[black] (43,1) --(43.5,1);

\node[scale=.8] at (46.9,1) {= 0};

\end{tikzpicture}

\end{center}
\caption{Type of UV divergent diagrams that mutually cancel at order $\mathcal{O}(\beta^4)$.}
\label{fig:cancellations}
\end{figure}
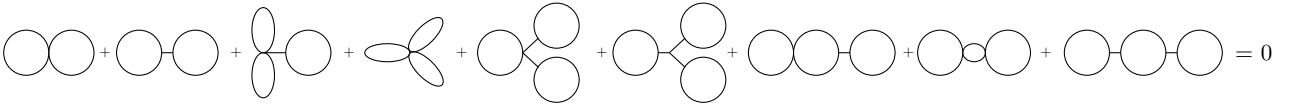

\textbf{UV divergences $\&$ cancellations.}
We summarise all the UV divergent diagrams. As expected these divergences appear logarithmically for large $\boldsymbol{l}$: 
\begin{align}
&\frac{1}{2}\sum_{\boldsymbol{l}\neq \bold{1}}\frac{(2l_1+1)(2l_2+1)}{A_{l_1}A_{l_2}}+\frac{1}{2}\sum_{\boldsymbol{l}\neq \bold{1}}\frac{(2l_1+1)(2l_2+1)}{A_{l_1}A_{l_2}}+ \frac{2}{A_0}\sum_{\boldsymbol{l}\neq \bold{1}}\frac{(2l_1+1)(2l_2+1)}{A_{l_1}A_{l_2}}\cr
&+ \frac{1}{A_0} \sum_{\boldsymbol{l}\neq \bold{1}} \frac{(2l_1+1)(2l_2+1)(2l_3+1)}{A_{l_1}A_{l_2}A_{l_3}}+\frac{1}{6} \sum_{\boldsymbol{l}\neq \bold{1}}\frac{(2l_1+1)(2l_2+1)(2l_3+1)}{A_{l_1}A_{l_2}A_{l_3}} + \frac{2}{A_0^2}\sum_{\boldsymbol{l}\neq \bold{1}}\frac{(2l_1+1)(2l_2+1)(2l_3+1)}{A_{l_1}A_{l_2}A_{l_3}}\cr
&+ \frac{4}{3A_{0}^3} \sum_{\boldsymbol{l}\neq \bold{1}}\frac{(2l_1+1)(2l_2+1)(2l_3+1)}{A_{l_1}A_{l_2}A_{l_3}}+\frac{1}{8} \sum_{\boldsymbol{l}\neq \bold{1}}\frac{(2l_1+1)(2l_2+1)(2l_3+1)(2l_4+1)}{A_{l_1}A_{l_2}A_{l_3}A_{l_4}}\cr
& +\frac{1}{2A_0^2}\sum_{\boldsymbol{l}\neq \bold{1}}\frac{(2l_1+1)(2l_2+1)(2l_3+1)(2l_4+1)}{A_{l_1}A_{l_2}A_{l_3}A_{l_4}}+\frac{1}{2A_0}\sum_{\boldsymbol{l}\neq \bold{1}}\frac{(2l_1+1)(2l_2+1)(2l_3+1)(2l_4+1)}{A_{l_1}A_{l_2}A_{l_3}A_{l_4}}\cr
&+ \frac{4}{A_0}\sum_{\boldsymbol{l}\neq \bold{1}}\frac{(2l_1+1)(2l_2+1)(2l_3+1)}{A_{l_1}^2A_{l_2}A_{l_3}}+ \sum_{\boldsymbol{l}\neq \bold{1}}\frac{(2l_1+1)(2l_2+1)(2l_3+1)}{A_{l_1}A_{l_2}A_{l_3}^2}+  \frac{4}{A_0^2}\sum_{\boldsymbol{l}\neq \bold{1}}\frac{(2l_1+1)(2l_2+1)(2l_3+1)}{A_{l_1}^2 A_{l_2}A_{l_3}}\cr
&=0~,
\end{align}
where in the last step we used that $A_0=-2$ at order $\mathcal{O}(1)$. Graphically this corresponds to the cancellations of (fig. \ref{fig:cancellations})

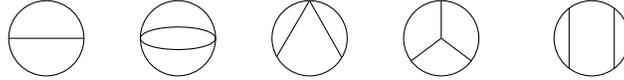
\begin{figure}[H]
\begin{center}
\begin{tikzpicture}[scale=.5]

\draw (-2.5,1) circle (1cm);
\draw[black] (-3.5,1) --(-1.5,1);

\draw (1,1) circle (1cm);
\draw (1,1) ellipse (1cm and .3cm);

\draw (4.5,1) circle (1cm);
\draw[black] (4.5,2) --(3.62,.5);
\draw[black] (4.5,2) --(5.35,.5);

\draw (8,1) circle (1cm);
\draw[black] (8,1) --(8,2);
\draw[black] (8,1) --(8.8,.4);
\draw[black] (8,1) --(7.2,.4);

\draw (12.,1) circle (1cm);
\draw[black] (12.6,1.8) --(12.6,.2);
\draw[black] (11.4,1.8) --(11.4,.2);

\end{tikzpicture}

\end{center}
\caption{Type of diagrams surviving at order $\mathcal{O}(\beta^4)$.}
\label{fig:diagramsBeta4_final}
\end{figure}

Additionally we observe cancellations between UV finite diagrams; graphically these are cancellations of diagrams of the form  (\ref{fig:finite_cancellations})
\begin{figure}[H]
\begin{center}
\begin{tikzpicture}[scale=.4]

\draw (-2.5,1) circle (1cm);
\draw (-.5,1) circle (1cm);
\draw[black] (-3.5,1) --(-1.5,1);

\node[scale=.5] at (1,1) {+};

\draw (2.5,1) circle (1cm);
\draw[black] (1.5,1) --(4.,1);
\draw (5,1) circle (1cm);

\node[scale=.5] at (6.5,1) {+};

\draw (8,1) circle (1cm);
\draw (10,1) circle (1cm);
\draw[black] (10,2) --(10,0);

\node[scale=.5] at (11.5,1) {+};

\draw (13,1) circle (1cm);
\draw (15.5,1) circle (1cm);
\draw[black] (14,1) --(14.5,1);
\draw[black] (15.5,2) --(15.5,0);

\node[scale=.8] at (17.7,1) {= 0};

\end{tikzpicture}

\end{center}
\caption{Type of UV finite diagrams that mutually cancel at order $\mathcal{O}(\beta^4)$.}
\label{fig:finite_cancellations}
\end{figure}
\noindent
\textbf{Finite $\mathcal{O}(\beta^4)$ diagrams.} We are finally left with the following 7 diagrams: 
\begin{align}\nonumber\label{3loop_diagrams}
&\text{loops}_{\beta^4}\equiv 4\sum_{\boldsymbol{l}\neq \textbf{1}} \frac{(2l_1+1)(2l_2+1)(2l_3+1)}{A_{l_1}^2A_{l_2} A_{l_3}} \begin{pmatrix} l_1 & l_2 & l_3 \\ 0 & 0 & 0\end{pmatrix}^2+\frac{4}{3}\sum_{\boldsymbol{l}\neq \bold{1}}\frac{(2l_1+1)(2l_2+1)(2l_3+1)}{A_{l_1}A_{l_2}A_{l_3}}\begin{pmatrix} l_1 & l_2 & l_3 \\ 0 & 0 & 0\end{pmatrix}^2
\\ \nonumber
&+\frac{1}{3}\sum_{\boldsymbol{l}\neq \bold{1}}\sum_{l_5\geq 0}\frac{(2l_1+1)(2l_2+1)(2l_3+1)(2l_4+1)(2l_5+1)}{A_{l_1}A_{l_2}A_{l_3}A_{l_4}}\begin{pmatrix} l_1 & l_2 & l_5\\ 0 & 0 & 0\end{pmatrix}^2\begin{pmatrix} l_3 & l_4 & l_5\\ 0 & 0 & 0\end{pmatrix}^2\\ \nonumber
&+4 \sum_{\boldsymbol{l}\neq \bold{1}} \frac{(2l_1+1)(2l_2+1)(2l_3+1)(2l_4+1)(2l_5+1)}{A_{l_1}A_{l_2}A_{l_3}A_{l_4}A_{l_5}}\begin{pmatrix}  l_1 & l_2 & l_5 \\ 0 & 0 & 0\end{pmatrix}^2 \begin{pmatrix}  l_3 & l_4 & l_5 \\ 0 & 0 & 0\end{pmatrix}^2\\ \nonumber
&+ \frac{2}{9}\sum_{\boldsymbol{l}\neq \bold{1}}\frac{(2l_1+1)(2l_2+1)(2l_3+1)(2l_4+1)(2l_5+1)(2l_6+1)}{A_{l_1}A_{l_2}A_{l_3}A_{l_4}A_{l_5}A_{l_6}}\begin{pmatrix} l_1 & l_2 & l_3 \\ 0 & 0 & 0\end{pmatrix}^2\begin{pmatrix} l_4 & l_5 & l_6 \\ 0 & 0 & 0\end{pmatrix}^2\\ \nonumber
&+4 \sum_{\boldsymbol{l}\neq \bold{1}}\frac{(2l_1+1)(2l_2+1)(2l_3+1)(2l_4+1)(2l_5+1)}{A_{l_1}A_{l_2}A_{l_3}A_{l_4}A_{l_5}^2}\begin{pmatrix}
l_1 & l_2 & l_5\\
0 & 0 & 0
\end{pmatrix}^2\begin{pmatrix}
l_3 & l_4 & l_5\\
0 & 0 & 0
\end{pmatrix}^2\cr
&+ \frac{8}{3} \sum_{\boldsymbol{l}\neq \bold{1}}(-1)^{m_1+m_2+m_3+m_4+m_5+m_6}\frac{(2l_1+1)(2l_2+1)(2l_3+1)(2l_4+1)(2l_5+1)(2l_6+1)}{A_{l_1}A_{l_2}A_{l_3}A_{l_4}A_{l_5}A_{l_6}}\\ 
&\times 
\begin{pmatrix}
l_1 & l_2 & l_3\\
m_1 & m_2 & m_3
\end{pmatrix}
\begin{pmatrix}
l_1 & l_4 & l_5\\
-m_1 & m_4 & m_5
\end{pmatrix}
\begin{pmatrix}
l_2 & l_4 & l_6\\
-m_2 & -m_4 & m_6
\end{pmatrix}
\begin{pmatrix}
l_3 & l_5 & l_6\\
-m_3 & -m_5 & -m_6
\end{pmatrix}\cr
&\times \begin{pmatrix}
l_1 & l_2 & l_3\\
0 & 0 & 0
\end{pmatrix}
\begin{pmatrix}
l_1 & l_4 & l_5\\
0 & 0 & 0
\end{pmatrix}
\begin{pmatrix}
l_2 & l_4 & l_6\\
0 & 0 & 0
\end{pmatrix}
\begin{pmatrix}
l_3 & l_5 & l_6\\
0 & 0 & 0
\end{pmatrix}
%&\times \sum_{m_1,m_2}(-1)^{m_1+m_2}\begin{pmatrix}  l_1 & l_2 & l_5 \\ m_1 & m_2 & -m_1-m_2\end{pmatrix}\begin{pmatrix}  l_1 & l_2 & l_5 \\ m_1 & m_2 & m_1+m_2\end{pmatrix}
+\text{two-loop}_{\beta^4}~,
\end{align}
where we denote by two-loop$_{\beta^4}$ the last three $\mathcal{O}(\beta^4)$ contributions of the two-loop diagrams in (\ref{eq:twoloop_hO}) (those proportional to $a_1/a_0$). We remark that the fifth diagram in the above contribution is minus one half times the square of the melonic diagram (\ref{melons}).
Out of all the 26 diagrams contributing to the path integral  (\ref{Zpert2}) at three-loop only the above seven UV finite diagrams survive. All the UV divergent diagrams cancel each other mutually because of the coefficients in (\ref{Zpert2}) as well as the fact that $A_0= -2$ (\ref{propagator}). The diagrams contributing to loop$_{\beta^4}$ are of the form depicted in the figure below (\ref{fig:diagramsBeta4_final}). 

\subsection{Three-loop two-sphere partition function}
Combining (\ref{ZLclas}), (\ref{Zpert22}), (\ref{FP_diagrams}) as well as loops$_{\beta^4}$, the semiclassical two-sphere partition function in TLT at three-loop order is given by 
\begin{align}\label{Ztl_3Loop}
&~\mathcal{Z}_{tL}[\Lambda] \approx
 \frac{\pm i }{\text{vol}_{SO(3)}} \, \mathrm{const}\,\times \,e^{-\frac{1}{\beta^2}- \frac{1}{\beta^2}\log\left(4\pi  \beta^2\right)} \upsilon^{\frac{c_{tL}}{6}}\Lambda_{\mathrm{uv}}^{\frac{7}{6}- \beta^2} \Lambda^{-\frac{1}{\beta^2}+1} \times \bigg(\frac{1}{\beta} \\
&+ \left(\frac{1}{6}(19-6\log 4)- (2\gamma_E+\log \pi)+ \left(\mathlarger{\mathlarger{\ominus}}- \frac{10}{3}+ \frac{5\pi a_1}{2a_0}\right)\right)\beta\cr
&+\Bigg(\frac{1}{2}\times \left(\frac{1}{6}(19-6\log 4)- (2\gamma_E+\log \pi)+ \left(\mathlarger{\mathlarger{\ominus}}- \frac{10}{3}+ \frac{5\pi a_1}{2a_0}\right)\right)^2 \cr
&+\left(\text{loops}_{\beta^4}-\frac{17}{27}+ \frac{15\pi a_1}{4a_0}-\frac{25\pi^2a_1^2}{8a_0^2}-\frac{1}{2}\mathlarger{\mathlarger{\ominus}}^2\right)\Bigg)\beta^3+...\bigg)~,\nonumber
%& - \left(\frac{1}{6}- \frac{5\pi}{2}\left(\frac{a_1}{a_0}\right)+2\gamma_E+ \log 4\pi-  \mathlarger{\mathlarger{\ominus}} \right)\beta 
% +\bigg(\frac{1}{2}\left(\frac{1}{6}- \frac{5\pi}{2}\left(\frac{a_1}{a_0}\right)+2\gamma_E+ \log 4\pi-  \mathlarger{\mathlarger{\ominus}} \right)^2 \cr
%&+\frac{10\pi}{3}\left(\frac{a_1}{a_0}\right)- \frac{17}{27}
% -\frac{1}{2}\mathlarger{\mathlarger{\ominus}}^2 +\left(\frac{1}{6}- \frac{5\pi}{2}\left(\frac{a_1}{a_0}\right)+2\gamma_E+ \log 4\pi\right)\mathlarger{\mathlarger{\ominus}}+ 3\text{-loop}\bigg)\beta^3+ ... \Big)~,
\end{align}
%\begin{align}\label{Ztl_3Loop}
%&~\mathcal{Z}_{tL}[\Lambda] \approx
% \frac{\pm i }{\text{vol}_{SO(3)}} \, \mathrm{const}\,\times \,e^{-\frac{1}{\beta^2}- \frac{1}{\beta^2}\log\left(4\pi  \beta^2\right)} \upsilon^{\frac{c_{tL}}{6}}\Lambda_{\mathrm{uv}}^{\frac{7}{6}- \beta^2} \Lambda^{-\frac{1}{\beta^2}+1} \times \bigg(\frac{1}{\beta} \\
%&+ \left(\frac{1}{6}(19-6\log 4)- (2\gamma_E+\log \pi)+ \left(\mathlarger{\mathlarger{\ominus}}- \frac{10}{3}+ \frac{5\pi a_1}{2a_0}\right)\right)\beta\cr
%&+\left(\frac{1}{2}\times \frac{1}{36}(19-6\log 4)^2+ \frac{1}{2}(2\gamma_E+\log \pi)^2+\left(\text{loops}_{\beta^4}-\frac{17}{27}+ \frac{15\pi a_1}{4a_0}-\frac{25\pi^2a_1^2}{8a_0^2}\right)\right)\beta^3+...\bigg)~,\nonumber
%%& - \left(\frac{1}{6}- \frac{5\pi}{2}\left(\frac{a_1}{a_0}\right)+2\gamma_E+ \log 4\pi-  \mathlarger{\mathlarger{\ominus}} \right)\beta 
%% +\bigg(\frac{1}{2}\left(\frac{1}{6}- \frac{5\pi}{2}\left(\frac{a_1}{a_0}\right)+2\gamma_E+ \log 4\pi-  \mathlarger{\mathlarger{\ominus}} \right)^2 \cr
%%&+\frac{10\pi}{3}\left(\frac{a_1}{a_0}\right)- \frac{17}{27}
%% -\frac{1}{2}\mathlarger{\mathlarger{\ominus}}^2 +\left(\frac{1}{6}- \frac{5\pi}{2}\left(\frac{a_1}{a_0}\right)+2\gamma_E+ \log 4\pi\right)\mathlarger{\mathlarger{\ominus}}+ 3\text{-loop}\bigg)\beta^3+ ... \Big)~,
%\end{align}
where 
\begin{equation}\label{A0}
\mathrm{const} \equiv a_0\times \frac{6  \times 2^{1/3} \sqrt{3} }{ A^2  \pi ^{5/6}} e^{-25/12}~,\quad \frac{a_1}{a_0}= \frac{27}{20\pi}~.
%\frac{3 \sqrt{3}\, e^{-\frac{25}{12}}}{16\times 2^{1/6}\,A^2\,\pi^{7/3}}~.
\end{equation}
and $c_{tL}=1-6q^2$ is given by
\begin{equation}
c_{tL}= -\frac{6}{\beta^2}+ 13-6\beta^2~.
\end{equation}
The partition function (\ref{Ztl_3Loop}) clearly reflects the sphere anomaly of a two-dimensional CFT, providing independent evidence for timelike Liouville to be a conformal field theory. 
Numerically we obtain strong evidence (see appendix \ref{appendix_Loops}) that at two-loop
\begin{equation}\label{conjecture_melons}
\mathlarger{\mathlarger{\ominus}} -\frac{10}{3}+\frac{5\pi a_1}{2a_0}=0~,\quad \frac{a_1}{a_0}= \frac{27}{20\pi}~,
\end{equation}
where we defined the melonic diagrams in (\ref{melons}). We thus end up with 
\begin{align}\label{Ztl_3Loop2}
&~\mathcal{Z}_{tL}[\Lambda] \approx
 \frac{\pm i }{\text{vol}_{SO(3)}} \, \mathrm{const}\,\times \,e^{-\frac{1}{\beta^2}- \frac{1}{\beta^2}\log\left(4\pi  \beta^2\right)} \upsilon^{\frac{c_{tL}}{6}}\Lambda_{\mathrm{uv}}^{\frac{7}{6}- \beta^2} \Lambda^{-\frac{1}{\beta^2}+1} \times \bigg(\frac{1}{\beta} \cr
&+ \left(\frac{1}{6}(19-6\log 4)- (2\gamma_E+\log \pi)\right)\beta+\Bigg(\frac{1}{2}\times \left(\frac{1}{6}(19-6\log 4)- (2\gamma_E+\log \pi)\right)^2 \cr
&+\left(\text{loops}_{\beta^4}-\frac{17}{27}+ \frac{15\pi a_1}{4a_0}-\frac{25\pi^2a_1^2}{8a_0^2}-\frac{1}{2}\mathlarger{\mathlarger{\ominus}}^2\right)\Bigg)\beta^3+...\bigg)~.
%& - \left(\frac{1}{6}- \frac{5\pi}{2}\left(\frac{a_1}{a_0}\right)+2\gamma_E+ \log 4\pi-  \mathlarger{\mathlarger{\ominus}} \right)\beta 
% +\bigg(\frac{1}{2}\left(\frac{1}{6}- \frac{5\pi}{2}\left(\frac{a_1}{a_0}\right)+2\gamma_E+ \log 4\pi-  \mathlarger{\mathlarger{\ominus}} \right)^2 \cr
%&+\frac{10\pi}{3}\left(\frac{a_1}{a_0}\right)- \frac{17}{27}
% -\frac{1}{2}\mathlarger{\mathlarger{\ominus}}^2 +\left(\frac{1}{6}- \frac{5\pi}{2}\left(\frac{a_1}{a_0}\right)+2\gamma_E+ \log 4\pi\right)\mathlarger{\mathlarger{\ominus}}+ 3\text{-loop}\bigg)\beta^3+ ... \Big)~,
\end{align}
We finish this section by summarising the main ingredients of (\ref{Ztl_3Loop2})
\begin{itemize}
\item The $\pm i$ ambiguity arises after Wick rotating the unsuppressed $l=0$ mode backwards.
\item The $\upsilon$ dependency is reminiscent of the sphere anomaly of a two-dimensional CFT providing independent evidence that TLT is indeed a two-dimensional CFT. 
\item The Fadeev-Popov determinant adds $\mathcal{O}(\beta^{-3})$ to the semiclassical expansion.
\end{itemize}
\section{Comparison to (timelike) DOZZ}\label{sec:4}
The goal of this section is to compare (\ref{Ztl_3Loop2}) to the sphere partition function obtained upon analytically continuing the DOZZ formula for three area operators $\mathcal{O}_\beta =e^{2\beta\varphi}$.
\subsection{Sphere partition function from (timelike) DOZZ}
Another possibility to obtain the sphere partition function exploits the DOZZ formula of three area operators. The three-point function structure constant has originally been introduced for spacelike Liouville theory. Whereas the analytic continuation $\varphi \rightarrow \pm i \varphi$, $b\rightarrow \mp i\beta$ and $Q\rightarrow \pm i q$ from the spacelike to the timelike Liouville action is an admissible process, it is in general not for the three-point function. Care must be taken because of the pole structure of the DOZZ formula. It seems however that the spacelike DOZZ formula for three area operators $\mathcal{O}_b = e^{2b\varphi}$ admits a well defined analytic continuation. Since additionally the timelike DOZZ formula in \cite{Harlow:2011ny} for three area operators $\mathcal{O}_\beta = e^{2\beta\varphi}$ vanishes, contradicting a non-vanishing sphere partition function from a path integral perspective we will proceed with our comparison using $C(b,b,b; \Lambda)|_{b\rightarrow \pm i\beta}$.
We thus consider
\begin{equation}
\langle \mathcal{O}_{\beta}(z_1) \mathcal{O}_{\beta}(z_2) \mathcal{O}_{\beta}(z_3) \rangle =  \frac{1}{\text{vol}_{PSL(2,\mathbb{C})}} \times \frac{C(b,b,b; \Lambda)|_{b\rightarrow \pm i\beta}}{|z_1-z_2|^2 |z_1-z_3|^2 |z_2-z_3|^2}~.
\end{equation}
To make contact with the Liouville partition function $\mathcal{Z}_{tL
}[\Lambda]$ in (\ref{ZL}) we note that \cite{Giribet:2011zx}
\begin{equation}
- \partial_\Lambda^3\mathcal{Z}_{tL}^{\text{DOZZ}}[\Lambda] = 2 \times C(b,b,b;\Lambda)|_{b\rightarrow \pm i\beta}~,
% \int_{\mathbb{C}^3} \frac{d^2 z_1 d^2 z_2 d^2 z_3}{|z_1-z_2|^2 |z_1-z_3|^2 |z_2-z_3|^2}~,
\end{equation}
where we have used that \cite{Zamolodchikov:1995aa}
\begin{equation}
\int_{\mathbb{C}^3} \frac{\dd^2 z_1 \dd^2 z_2 \dd^2 z_3}{|z_1-z_2|^2 |z_1-z_3|^2 |z_2-z_3|^2} =  2 \, {\text{vol}_{PSL(2,\mathbb{C})}}~.
\end{equation}
This leads to \cite{timelike,Giribet:2011zx}
\begin{equation}\label{tLZs2}
\mathcal{Z}^{\text{DOZZ}}_{tL}[\Lambda] = \pm i\left(\pi \Lambda \gamma(-\beta^2)\right)^{-\frac{1}{\beta^2}+1}\frac{(1+\beta^2)}{\pi^3 q \gamma(-\beta^2)\gamma(-\beta^{-2})}\,e^{q^2 -q^2 \log 4}~,
\end{equation}
where $\gamma(x) \equiv \Gamma(x)/\Gamma(1-x)$.
In the semiclassical $\beta \rightarrow 0^+$ limit we obtain 
\begin{align}\label{eq:sphere_DOZZ}
\mathcal{Z}^{\text{DOZZ}}_{tL}[\Lambda]
% &= \pm e^{-\frac{1}{\beta^2}- \frac{1}{\beta^2}\log (4\pi \beta^2) }\Lambda^{-\frac{1}{\beta^2}+1} \left( 1 - e^{\frac{2i\pi}{\beta^2}}\right) \cr
%&\times \frac{1}{\beta} e^{(1-\log 4)\beta^2-2\sum_{n=1}^\infty \frac{\zeta(2n+1)}{2n+1}\beta^{4n} +2 \beta^2\sum_{n=1}^\infty \frac{1}{2n+1}\beta^{4n}+\frac{1}{6}\beta^2-\frac{1}{180}\beta^6+ \frac{1}{630}\beta^{10} -\frac{1}{840}\beta^{14}+ \frac{1}{594}\beta^{18} - \frac{691}{180180}\beta^{22}+ \frac{1}{30}\beta^{26}+ \ldots}\cr
% &= \pm e^{-\frac{1}{\beta^2}- \frac{1}{\beta^2}\log (4\pi \beta^2) }\Lambda^{-\frac{1}{\beta^2}+1} \left( 1 - e^{\frac{2i\pi}{\beta^2}}\right) \cr
%&\times \frac{1}{\beta} e^{(1-\log 4)\beta^2-2\sum_{n=1}^\infty \frac{\zeta(2n+1)}{(2n+1)}\beta^{4n}+ \frac{13}{6}\beta^2+ \frac{119}{180}\beta^6+ \frac{253}{630}\beta^{10}+ \frac{239}{840}\beta^{14}+ \frac{133}{594}\beta^{18}+\frac{32069}{180180}\beta^{22}+ \frac{1}{6}\beta^{26} \ldots }
%\cr
& \approx \pm \frac{16}{\pi^2}e^{-2-2\gamma_E}e^{-\frac{1}{\beta^2}- \frac{1}{\beta^2}\log (4\pi \beta^2) }\Lambda^{-\frac{1}{\beta^2}+1} \left( 1 - e^{\frac{2i\pi}{\beta^2}}\right)\cr
&\times \left(\frac{1}{\beta}+ {\frac{1}{6}\left({19}- 6\log 4\right)}\beta+ \left(\frac{1}{2}\times \frac{1}{36} (19-6 \log 4)^2-\frac{2}{3} \zeta (3)\right)\beta^3+\ldots \right)~,
\end{align}
where we have taken $e^{-i\pi}=-1$. This we compare with (\ref{Ztl_3Loop}). However we quickly realise that (\ref{eq:sphere_DOZZ}) bears two subtleties, which we can phrase in terms of two questions:
\begin{itemize}
\item What is the regularisation scheme of the DOZZ formula?
\item What is the meaning of $e^{\frac{2i\pi}{\beta^2}}$ in $\mathcal{Z}^{\text{DOZZ}}_{tL}[\Lambda] $?
\end{itemize}
To attack the first question we allow some freedom in the UV cutoff in (\ref{Ztl_3Loop}). In our path integral derivation we chose a specific regularisation scheme---a heat kernel approach---to calculate the functional determinant (\ref{heatkernel}). To compare the two-sphere partition functions we thus need to allow for an additional parameter $\Lambda_{\text{uv}}\rightarrow s\Lambda_{\text{uv}}$ keeping track of this scheme. Motivated by a two-loop comparison of (\ref{Ztl_3Loop}) and (\ref{eq:sphere_DOZZ}) and using (\ref{conjecture_melons}) we choose\footnote{Note that this is slightly different than $s$ in v1 and v2 of \cite{timelike} where we did not Wick rotate the Fadeev-Popov determinant. The comparison with the DOZZ formula at three-loop however suggests that we have to Wick rotate $\varphi_{2,m}$ in $\Delta_{\text{FP}}$ rendering $a_1/a_0$ positive.} 
\begin{equation}\label{s}
s= e^{-\left(2\gamma_E+\log\pi\right)}~.
\end{equation}
It remains to test whether
\begin{align}\label{finaltest}
0&\overset{?}{=} -\frac{25\pi^2}{8}\left(\frac{a_1}{a_0}\right)^2+\frac{15\pi}{4}\left(\frac{a_1}{a_0}\right)- \frac{17}{27}+{\text{loops}}_{\beta^4}-\frac{1}{2}\mathlarger{\mathlarger{\ominus}}^2+\frac{2}{3}\zeta(3) ~,
\end{align}
holds true. This equation combines contributions arising from the Fadeev-Popov gauge fixing procedure, apparent in the ratio $a_1/a_0$ (\ref{a1a0}), the functional determinant analysis and the loop expansion. It is clear that the challenge consists in evaluating the diagrams in (\ref{3loop_diagrams}), in particular the last contributions in loops$_{\beta^4}$ stemming from 
\begin{equation}
\bigg\langle
\int_{S^2} \dd\Omega \dd\Omega'\dd\Omega''\dd{\Omega'''} \varphi(\Omega)^3\varphi(\Omega')^3\varphi(\Omega'')^3\varphi(\Omega''')^3\bigg\rangle~.
\end{equation}
We rely on a numerical evaluation of the $3j$ symbols in  (\ref{3loop_diagrams}) closely approaching the exact result (\ref{finaltest}) (see appendix \ref{appendix_Loops} for some numerical evidence). We believe that proving (\ref{finaltest}) exactly is solely a numerical issue.\footnote{In case numerical explorations would not confirm (\ref{finaltest}) another possible option to explore would be a $\beta$ dependent shift $s\rightarrow s(\beta)$.}
%Numerically we find good evidence for this relation. It thus allows us to conjecture the all loop structure
%\begin{align}
%0&\overset{?}{=} \frac{125 \pi ^3}{24}\left(\frac{a_1}{a_0}\right)^3-\frac{75 \pi ^2}{8 }\left(\frac{a_1}{a_0}\right)^2+\frac{45 \pi }{8 }\left(\frac{a_1}{a_0}\right)-\frac{41}{48}+{\text{loops}}_{\beta^6}-\frac{119}{180}~,\cr
%0&\overset{?}{=} -\frac{625 \pi ^4 }{64}\left(\frac{a_1}{a_0}\right)^4+\frac{375 \pi ^3 }{16 }\left(\frac{a_1}{a_0}\right)^3-\frac{675 \pi ^2 }{32 }\left(\frac{a_1}{a_0}\right)^2+\frac{115 \pi }{16}\left(\frac{a_1}{a_0}\right)-\frac{343}{640}+\text{loops}_{\beta^8}- \frac{2}{7}\zeta(7)~,\cr
%0&\overset{?}{=}\frac{625 \pi ^5}{32 }\left(\frac{a_1}{a_0}\right)^5-\frac{1875 \pi ^4 }{32}\left(\frac{a_1}{a_0}\right)^4+\frac{1125 \pi ^3}{16 }\left(\frac{a_1}{a_0}\right)^3-\frac{625 \pi ^2
% }{16}\left(\frac{a_1}{a_0}\right)^2+\frac{285 \pi  }{32 }\left(\frac{a_1}{a_0}\right)-\frac{101}{320}\cr
% &+{\text{loops}}_{\beta^{10}}- \frac{253}{630}~.
%\end{align}

To approach the second question we allow ourselves a short detour. One possible interpretation of the exponential in $e^{\frac{2i\pi}{\beta^2}}$ is as a second \textit{complex} saddle in the path integral with $\varphi_c= \varphi_*+ \pi i/\beta$. Whereas in principle the gravitational path integral allows for the integer indexed family of complex constant solutions $\varphi_{c,*}= \varphi_{*}+ \pi i n/\beta, n\in \mathbb{Z}$ (\ref{eoms}) the DOZZ formula dictates that only the complex saddle with $n=1$ should be included in the path integral. It is not clear to us why this specific saddle needs to be included. Since (\ref{metric_start})
\begin{equation}
\dd s^2 = e^{2\beta \varphi_c}\dd\tilde{s}^2= e^{2\beta \varphi_*+ 2\pi i}\dd \tilde{s}^2= e^{2\beta \varphi_*}\dd \tilde{s}^2~,
\end{equation}
it does not affect the metric and hence following \cite{Witten:2021nzp} both saddles are allowed. 

Finally we note that the agreement between (\ref{eq:sphere_DOZZ}) and (\ref{Ztl_3Loop}) provides further evidence that the semiclassical expansion provides a loophole around Gribov-phenomena \cite{Gribov:1977wm}. 

\section{Outlook}\label{sec:5}

\begin{figure}[H]
\begin{center}
\begin{tikzpicture}[scale=.5]

\draw (-5.5,1) circle (1cm);
\draw[black] (-6,0.12) --(-6,1.88);
\draw[black] (-5.,0.12) --(-5.,1.88);
\draw[black] (-5,1) --(-6,1);

\draw (-2.5,1) circle (1cm);
\draw[black] (-3.5,1) --(-1.5,1);
\draw[black] (-2.5,2) --(-2.5,0);

\draw (1,1) circle (1cm);
\draw (1,1.3) ellipse (.96cm and .15cm);
\draw (1,.7) ellipse (.96cm and .15cm);

\draw (4.5,1) circle (1cm);
\draw[black] (4.5,2) --(5.5,1);
\draw[black] (4.5,2) --(3.5,1);
\draw[black] (4.5,0) --(3.5,1);
\draw[black] (4.5,0) --(5.5,1);

\draw (8,1) circle (1cm);
\draw (8,1) ellipse (1cm and .6cm);
\draw (8,1) ellipse (1cm and .15cm);

\draw (11.5,1) circle (1cm);
\draw[black] (12.1,1.8) --(12.1,.2);
\draw[black] (10.9,1.8) --(10.9,.2);
\draw[black] (11.5,0) --(11.5,2);

\end{tikzpicture}

\end{center}
\caption{Examples of diagrams expected to survive cancellations at four- and five-loop.}
\label{fig:fourfive}
\end{figure}
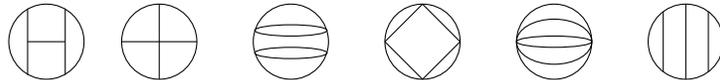

\textbf{All-loops path integral.} The analytically continued DOZZ formula provides the all-loop conjectured sphere partition function of two-dimensional quantum gravity. Provided we take into consideration the scheme dependency of the regularisation scheme (\ref{s}) its semiclassical expansion agrees with the path integral expansion. Importantly however the DOZZ formula predicts the inclusion of one additional complex saddle in the path integral picture. It would be interesting to understand why the DOZZ formula picks one and not multiple additional complex saddles. Furthermore we observe that at two- and three-loop all UV divergences in the path integral cancel and that the surviving diagrams are some sort of generalised melonic type diagrams (fig. \ref{fig:fourfive}).  It would be interesting to understand if this is the general structure of diagrams in TLT on the two-sphere. Besides being non-unitary timelike Liouville theory provides a well defined UV finite QFT on the sphere. It would be interesting to understand if there is a symmetry argument that could explain the cancellations of the UV divergent \textit{and} UV finite diagrams observed in loop calculations. Maybe understanding the set-up in position space could be helpful for this also.
\noindent
\textbf{Higher genera.} Whereas timelike Liouville theory provides an unconstrained sphere saddle on a genus zero surface, as compared to (spacelike) Liouville theory which does so only upon fixing the area of the physical metric \cite{Zamolodvarphikov:1982vx,beatrix}, upon increasing the genus of the Riemann surface spacelike and TLT switch roles. Since the Ricci scalar is negative for $h\geq 2$ TLT now only admits a saddle upon fixing the area of the physical metric.  
\newline
\textbf{A microscopic model for timelike Liouville?} Although various attempts of a microscopic model for timelike Liouville theory have been discussed (see e.g. \cite{Takayanagi:2004yr}) none of them has so far been able to provide sufficient evidence. It would be interesting to see whether the comparison of the sphere partition function of timelike Liouville theory with the sphere partition function of spacelike Liouville theory \cite{beatrix,Zamolodvarphikov:1982vx} can provide some evidence for a microscopic model. This model might be more subtle than a matrix model. 
\newline
\textbf{Supersymmetric timelike Liouville theory.} 
In \cite{Rashkov:1996np,Poghossian:1996agj} a supersymmetric version of spacelike Liouville theory has been introduced. It would be interesting to extend this to timelike supersymmetric Liouville theory. In particular it would be interesting to understand if $\mathcal{N}=1$ timelike super Liouville admits a dS$_2$ saddle and could thus provide a setup to use supersymmetric techniques in de Sitter \cite{Anninos:2021ihe, DDB} 

%\textbf{Inconsistency.} The major open question remains to understand the miss-match in (\ref{finaltest}). Besides the possibility of an error in my calculations a possibility is a. Gribov-type effect \cite{Gribov:1977wm}. Maybe the semiclassical expansion fails at this order. Since from an easy argument of the 3j coefficients the summands of the diagrams in (\ref{3loop_diagrams}) fall off rather rapidly the major contribution in  (\ref{finaltest}) comes from the first three term. A different gauge choice is discussed in \cite{Maltz:2012zs}. It would be interesting to explore if this could lift the inconsistency. \newline\newline
%\textbf{UV cancellations.} In section \ref{sec:3} we calculated the three-loop order $\mathcal{O}(\beta^4)$ contributions to (\ref{Zpert2}). Remarkably, all the UV divergences cancel out and actually more is true. The UV divergent diagrams completely cancel each other. This is a pretty remarkable result; usually these type of cancellations happen in supersymmetric theories, here it is an artefact of the coefficients in (\ref{Zpert2}) and the form of the propagator (\ref{divloop}). Timelike Liouville theory therefore constitutes up to three-loop a UV finite quantum field theory on the sphere. It would be interesting to show this at all orders in the small $\beta$ expansion.

\section*{Acknowledegements}
It is a great pleasure to acknowledge Dio Anninos, Simon Caron-Huot, Lorenz Eberhardt and John Stout for useful discussions. I am particularly thankful to Dio Anninos, Shreya Vardhan and Antonio Rotundo for useful comments on the draft. B.M. is supported in part by the Simons Foundation Grant No. 385602 and the Natural Sciences and Engineering Research Council of Canada (NSERC), funding reference number SAPIN/00047-2020.

\appendix

\section{Spherical harmonics}\label{Yapp}
We use real valued spherical harmonics throughout the paper. We denote by $\mathcal{Y}_{l m}(\theta,\phi)$ the complex spherical harmonics defined by 
\begin{equation}
\mathcal{Y}_{l m}(\theta,\phi)= \sqrt{\frac{(2l+1)}{4\pi}\frac{(l-m)!}{(l+m)!}}\, P_{l,m}(\cos\theta)\, e^{im\phi}~,
\end{equation}
where $P_{l,m}$ is the associated Legendre function, and $m \in [-l,l]$ with $l\in\mathbb{N}$. Real spherical harmonics $Y_{l m}(\theta,\phi)$ can be obtained using the linear combinations
\begin{equation}\label{real_complex}
{Y}_{l m}(\theta,\phi)= \begin{cases}
\frac{i}{\sqrt{2}}\left(\mathcal{Y}_{l m}(\theta,\phi)- (-1)^m \mathcal{Y}_{l, -m}(\theta,\phi)\right)~,\quad \mathrm{if}~ m<0\\
\mathcal{Y}_{l 0}(\theta,\phi)\\
\frac{1}{\sqrt{2}}\left(\mathcal{Y}_{l, -m}(\theta,\phi)+ (-1)^m \mathcal{Y}_{l m}(\theta,\phi)\right)~,\quad \mathrm{if}~ m>0~.
\end{cases}
\end{equation}
The Wigner 3j symbol is given by the Clebsch-Gordan coefficients and gives the integral of the product of three complex spherical harmonics
\begin{align}
&\int_{S^2} \dd\phi \dd\theta\sin\theta\,\mathcal{Y}_{l_1,m_1}(\theta,\phi)\mathcal{Y}_{l_2,m_2}(\theta,\phi)\mathcal{Y}_{l_3,m_3}(\theta,\phi)\cr
&= \sqrt{\frac{(2l_1+1)(2l_2+1)(2l_3+1)}{4\pi}}\begin{pmatrix} l_1 & l_2 & l_3 \\ 0 & 0 & 0\end{pmatrix}\begin{pmatrix} l_1 & l_2 & l_3 \\ m_1 & m_2 & m_3\end{pmatrix}~.
\end{align}
\textbf{3j symbol relations.}
The Clebsch-Gordan coefficients satisfy various properties. In particular they obey the orthogonality relation
\begin{equation}\label{orthoCG}
\sum_{\alpha,\beta}\begin{pmatrix} a & b & c \\ \alpha & \beta & \gamma\end{pmatrix}\begin{pmatrix} a & b & c' \\ \alpha & \beta & \gamma'\end{pmatrix}= \frac{1}{2c+1}\delta_{cc'}\delta_{\gamma\gamma'}~.
\end{equation}
Furthermore
\begin{equation}\label{reduced_CG}
\begin{pmatrix}
a & b & c \\ 
0 & 0 & 0 
\end{pmatrix} \neq 0\quad  \mathrm{iff} ~a+ b+ c \in 2\mathbb{Z}~\quad \&\quad \begin{pmatrix}
a & b & 0 \\ 
\alpha & \beta & 0 
\end{pmatrix}= \frac{(-1)^{a-\alpha}}{\sqrt{2a+1}}\delta_{ab}\delta_{\alpha-\beta}~.
\end{equation}
The 3j symbol is non-vanishing iff 
\begin{equation}
\alpha+\beta+\gamma=0~,\quad |a-b| \leq c < a+b~.
\end{equation}
The latter condition we refer to as the triangle condition. 
\section{More three-loops}\label{appendix_Loops}

The last two expectation values in (\ref{3loop_diagrams1}) involve ten and twelve fields respectively, allowing seven and eight possible Gaussian integral combinations respectively. We start with the integral over $\varphi(\Omega)^3\varphi(\Omega')^3\varphi(\Omega'')^4$:
\begin{align} \nonumber
&\frac{1}{108\pi^3}\bigg\langle\int_{S^2} \dd\Omega \dd\Omega'\dd\Omega'' \varphi(\Omega)^3\varphi(\Omega')^3\varphi(\Omega'')^4\bigg\rangle = \frac{1}{108\pi^3} {\color{black}c_1} \times \int_{S^2} \dd\Omega\dd\Omega' \dd\Omega'' G(\Omega,\Omega)G(\Omega,\Omega')G(\Omega',\Omega') G(\Omega'',\Omega'')^2\\ \nonumber
&+ \frac{1}{108\pi^3} {\color{black}c_2}\times \int_{S^2} \dd\Omega\dd\Omega' \dd\Omega'' G(\Omega,\Omega')^3G(\Omega'',\Omega'')^2+   \frac{1}{108\pi^3} {\color{black}c_3}\times \int_{S^2} \dd\Omega\dd\Omega' \dd\Omega'' G(\Omega,\Omega'')^3 G(\Omega',\Omega'')G(\Omega',\Omega')\\ \nonumber
&+\frac{1}{108\pi^3}{\color{black}c_4}\times \int_{S^2} \dd\Omega\dd\Omega'\dd\Omega'' G(\Omega,\Omega)G(\Omega',\Omega')G(\Omega'',\Omega'')G(\Omega,\Omega'')G(\Omega',\Omega'') \\ \nonumber
&+ \frac{1}{108\pi^3}{\color{black}c_5}\times \int_{S^2} \dd \Omega\dd\Omega'\dd\Omega'' G(\Omega,\Omega')^2 G(\Omega,\Omega'')G(\Omega',\Omega'')G(\Omega'',\Omega'')\\ \nonumber
&+ \frac{1}{108\pi^3}{\color{black}c_6}\times  \int_{S^2} \dd\Omega\dd\Omega'\dd\Omega'' G(\Omega,\Omega')G(\Omega',\Omega'')^2G(\Omega'',\Omega'')G(\Omega,\Omega)\\ 
&+ \frac{1}{108\pi^3}{\color{black}c_7}\times \int_{S^2} \dd\Omega\dd\Omega'\dd\Omega'' G(\Omega,\Omega')G(\Omega,\Omega'')^2G(\Omega',\Omega'')^2~,
\end{align}
where ${\color{black}c_1}=27$, ${\color{black}c_2}=18$, ${\color{black}c_3}=144$, ${\color{black}c_4}=108$ and ${\color{black}c_5}=216$, ${\color{black}c_6}= 216$, ${\color{black}c_7}=216$.
After the dust settles we find
\begin{align}\label{eq:phi32phi4}
&\frac{1}{108\pi^3}\bigg\langle\int_{S^2} \dd\Omega \dd\Omega'\dd\Omega'' \varphi(\Omega)^3\varphi(\Omega')^3\varphi(\Omega'')^4\bigg\rangle= \frac{1}{2A_0}\sum_{\boldsymbol{l}\neq \bold{1}}\frac{(2l_1+1)(2l_2+1)(2l_3+1)(2l_4+1)}{A_{l_1}A_{l_2}A_{l_3}A_{l_4}}\cr
&+\frac{1}{3} \sum_{\boldsymbol{l}\neq \bold{1}}\frac{(2l_1+1)(2l_2+1)(2l_3+1)(2l_4+1)(2l_5+1)}{A_{l_1}A_{l_2}A_{l_3}A_{l_4}A_{l_5}}\begin{pmatrix} l_1 & l_2 & l_3 \\ 0 & 0 & 0\end{pmatrix}^2\cr
&+ \frac{8}{3A_0} \sum_{\boldsymbol{l}\neq \bold{1}}\frac{(2l_1+1)(2l_2+1)(2l_3+1)(2l_4+1)}{A_{l_1}A_{l_2}A_{l_3}A_{l_4}}\begin{pmatrix} l_1 & l_2 & l_3 \\ 0 & 0 & 0\end{pmatrix}^2\cr
&+ \frac{2}{A_0^2}\sum_{\boldsymbol{l}\neq \bold{1}}\frac{(2l_1+1)(2l_2+1)(2l_3+1)}{A_{l_1}A_{l_2}A_{l_3}}+ 4 \sum_{\boldsymbol{l}\neq \bold{1}}\frac{(2l_1+1)(2l_2+1)(2l_3+1)(2l_4+1)}{A_{l_1}A_{l_2}A_{l_3}^2A_{l_4}}\begin{pmatrix} l_1 & l_2 & l_3 \\ 0 & 0 & 0\end{pmatrix}^2\cr
&+ \frac{4}{A_0}\sum_{\boldsymbol{l}\neq \bold{1}}\frac{(2l_1+1)(2l_2+1)(2l_3+1)}{A_{l_1}^2A_{l_2}A_{l_3}}\cr
&+ 4 \sum_{\boldsymbol{l}\neq \bold{1}} \frac{(2l_1+1)(2l_2+1)(2l_3+1)(2l_4+1)(2l_5+1)}{A_{l_1}A_{l_2}A_{l_3}A_{l_4}A_{l_5}}\begin{pmatrix}  l_1 & l_2 & l_5 \\ 0 & 0 & 0\end{pmatrix}^2 \begin{pmatrix}  l_3 & l_4 & l_5 \\ 0 & 0 & 0\end{pmatrix}^2~.
\end{align}
\begin{figure}[H]
\begin{center}
\begin{tikzpicture}[scale=.35]

\draw (-13,0) circle (1cm);
\draw[black] (-12,0) --(-11.5,0);
\draw (-10.5,0) circle (1cm);

\draw (-8,0) circle (1cm);
\draw (-6,0) circle (1cm);

\draw (-2.5,0) circle (1cm);
\draw[black] (-3.5,0) --(-1.5,0);

\draw (0,0) circle (1cm);
\draw (2,0) circle (1cm);

\draw (5.5,0) circle (1cm);
\draw[black] (4.5,0) --(7,0);
\draw (8,0) circle (1cm);

\draw (11.5,0) circle (1cm);
\draw[black] (12.5,0) --(13.5,.8);
\draw[black] (12.5,0) --(13.5,-.8);
\draw (14.3,1.4) circle (1cm);
\draw (14.3,-1.4) circle (1cm);

 \draw (19.5,0) circle (1cm);
  \draw (17.5,0) circle (1cm);
\draw[black] (17.5,-1) --(17.5,1);

\draw (22.5,0) circle (1cm);
\draw (24.5,0) circle (1cm);
\draw[black] (25.5,0) --(26,0);
\draw (27,0) circle (1cm);

\draw (30.5,0) circle (1cm);
\draw[black] (30.5,1) --(29.62,-.5);
\draw[black] (30.5,1) --(31.35,-.5);

%\draw (-11.5,0) circle (1cm);
%\draw[black] (-10.5,0) --(-9.5,.8);
%\draw[black] (-10.5,0) --(-9.5,-.8);
%\draw (-8.7,1.4) circle (1cm);
%\draw (-8.7,-1.4) circle (1cm);

% \draw (-5,0) circle (1cm);
%  \draw (-3,0) circle (1cm);
%\draw[black] (-3,-1) --(-3,1);
%
%
%\draw (0.5,0) circle (1cm);
%\draw (2.5,0) circle (1cm);
%\draw[black] (3.5,0) --(4,0);
%\draw (5,0) circle (1cm);
%
%\draw (8.5,0) circle (1cm);
%\draw[black] (8.5,1) --(7.62,-.5);
%\draw[black] (8.5,1) --(9.35,-.5);

\end{tikzpicture}
\end{center}
\caption{Diagrams appearing in the Wick contraction of $ \varphi(\Omega)^3\varphi(\Omega')^3\varphi(\Omega'')^4$.}
\label{fig:diagramsValpha}
\end{figure}
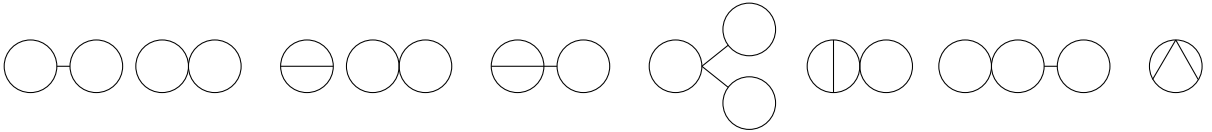
\noindent
Finally we have the last term in (\ref{3loop_diagrams}). We have twelve distinct fields leading to eight distinct non-vanishing Gaussian integrals
\begin{align}
&\frac{1}{1944\pi^4}\bigg\langle\int_{S^2} \dd\Omega \dd\Omega'\dd\Omega''\dd{\Omega'''} \varphi(\Omega)^3\varphi(\Omega')^3\varphi(\Omega'')^3\varphi(\Omega''')^3\bigg\rangle \cr
&=\frac{1}{1944\pi^4} {\color{black}c_1} \times \int_{S^2} \dd\Omega \dd\Omega'\dd\Omega''\dd{\Omega'''} G(\Omega,\Omega)G(\Omega,\Omega')G(\Omega',\Omega')G(\Omega'',\Omega'')G(\Omega'',\Omega''')G(\Omega''',\Omega''')\cr
&+ \frac{1}{1944\pi^4} {\color{black}c_2}\times  \int_{S^2} \dd\Omega \dd\Omega'\dd\Omega''\dd{\Omega'''}  G(\Omega,\Omega')^3 G(\Omega'',\Omega''')^3\cr
&+ \frac{1}{1944\pi^4} {\color{black}c_3}\times  \int_{S^2} \dd\Omega \dd\Omega'\dd\Omega''\dd{\Omega'''}  G(\Omega,\Omega)G(\Omega,\Omega')G(\Omega',\Omega')G(\Omega'',\Omega''')^3\cr
&+ \frac{1}{1944\pi^4} {\color{black}c_4}\times  \int_{S^2} \dd\Omega \dd\Omega'\dd\Omega''\dd{\Omega'''} G(\Omega,\Omega') G(\Omega,\Omega'')G(\Omega,\Omega''')G(\Omega',\Omega')G(\Omega'',\Omega'')G(\Omega''',\Omega''')\cr
&+ \frac{1}{1944\pi^4} {\color{black}c_5}\times  \int_{S^2} \dd\Omega \dd\Omega'\dd\Omega''\dd{\Omega'''} G(\Omega,\Omega') G(\Omega,\Omega'')G(\Omega,\Omega''')G(\Omega',\Omega'')^2 G(\Omega''',\Omega''')\cr
&+ \frac{1}{1944\pi^4} {\color{black}c_6}\times  \int_{S^2} \dd\Omega \dd\Omega'\dd\Omega''\dd{\Omega'''} G(\Omega,\Omega') G(\Omega,\Omega'')G(\Omega,\Omega''')G(\Omega',\Omega'')G(\Omega',\Omega''')G(\Omega'',\Omega''')\cr
&+ \frac{1}{1944\pi^4} {\color{black}c_7}\times  \int_{S^2} \dd\Omega \dd\Omega'\dd\Omega''\dd{\Omega'''} G(\Omega,\Omega')^2 G(\Omega'',\Omega''')^2 G(\Omega,\Omega'')G(\Omega',\Omega''')\cr
&+ \frac{1}{1944\pi^4} {\color{black}c_8}\times  \int_{S^2} \dd\Omega \dd\Omega'\dd\Omega''\dd{\Omega'''}  G(\Omega,\Omega'')^2 G(\Omega,\Omega')G(\Omega'',\Omega''')G(\Omega',\Omega')G(\Omega''',\Omega''')~,
\end{align}
where ${\color{black}c_1}= 3^4 \times 3$, ${\color{black}c_2}= 108$, ${\color{black}c_3}= 324$, ${\color{black}c_4}= 648$, ${\color{black}c_5}= 3888$, ${\color{black}c_8}= 1944$, ${\color{black}c_6}= 648 \times 2$, ${\color{black}c_7}=1944$.
After the dust settles we find 
\begin{align}\label{eq:phi34}
&\frac{1}{1944\pi^4}\bigg\langle\int_{S^2} \dd\Omega \dd\Omega'\dd\Omega''\dd{\Omega'''} \varphi(\Omega)^3\varphi(\Omega')^3\varphi(\Omega'')^3\varphi(\Omega''')^3\bigg\rangle \cr
&=  \frac{1}{2A_0^2}\sum_{\boldsymbol{l}\neq \bold{1}}\frac{(2l_1+1)(2l_2+1)(2l_3+1)(2l_4+1)}{A_{l_1}A_{l_2}A_{l_3}A_{l_4}}\cr
&+\frac{2}{9} \sum_{\boldsymbol{l}\neq \bold{1}}\frac{(2l_1+1)(2l_2+1)(2l_3+1)(2l_4+1)(2l_5+1)(2l_6+1)}{A_{l_1}A_{l_2}A_{l_3}A_{l_4}A_{l_5}A_{l_6}}\begin{pmatrix} l_1 & l_2 & l_3 \\ 0 & 0 & 0\end{pmatrix}^2\begin{pmatrix} l_4 & l_5 & l_6 \\ 0 & 0 & 0\end{pmatrix}^2\cr
&+\frac{2}{3A_0}\sum_{\boldsymbol{l}\neq \bold{1}}\frac{(2l_1+1)(2l_2+1)(2l_3+1)(2l_4+1)(2l_5+1)}{A_{l_1}A_{l_2}A_{l_3}A_{l_4}A_{l_5}}\begin{pmatrix} l_1 & l_2 & l_3 \\ 0 & 0 & 0\end{pmatrix}^2\cr
&+ \frac{4}{3A_{0}^3} \sum_{\boldsymbol{l}\neq \bold{1}}\frac{(2l_1+1)(2l_2+1)(2l_3+1)}{A_{l_1}A_{l_2}A_{l_3}}+  \frac{8}{A_0}\sum_{\boldsymbol{l}\neq \bold{1}}\frac{(2l_1+1)(2l_2+1)(2l_3+1)(2l_4+1)}{A_{l_1}^2A_{l_2}A_{l_3}A_{l_4}}\begin{pmatrix}
l_1 & l_2 & l_3\\
0 & 0 & 0 
\end{pmatrix}^2 \cr
&+ \frac{8}{3}\sum_{\boldsymbol{l}\neq \bold{1}}(-1)^{m_1+m_2+m_3+m_4+m_5+m_6}\frac{(2l_1+1)(2l_2+1)(2l_3+1)(2l_4+1)(2l_5+1)(2l_6+1)}{A_{l_1}A_{l_2}A_{l_3}A_{l_4}A_{l_5}A_{l_6}}\cr
&\times \begin{pmatrix}
l_1 & l_2 & l_3\\
0 & 0 & 0
\end{pmatrix}
\begin{pmatrix}
l_1 & l_4 & l_5\\
0 & 0 & 0
\end{pmatrix}
\begin{pmatrix}
l_2 & l_4 & l_6\\
0 & 0 & 0
\end{pmatrix}
\begin{pmatrix}
l_3 & l_5 & l_6\\
0 & 0 & 0
\end{pmatrix} \cr
&\times 
\begin{pmatrix}
l_1 & l_2 & l_3\\
m_1 & m_2 & m_3
\end{pmatrix}
\begin{pmatrix}
l_1 & l_4 & l_5\\
-m_1 & m_4 & m_5
\end{pmatrix}
\begin{pmatrix}
l_2 & l_4 & l_6\\
-m_2 & -m_4 & m_6
\end{pmatrix}
\begin{pmatrix}
l_3 & l_5 & l_6\\
-m_3 & -m_5 & -m_6
\end{pmatrix}\cr
&+ 4\sum_{\boldsymbol{l}\neq \bold{1}}\frac{(2l_1+1)(2l_2+1)(2l_3+1)(2l_4+1)(2l_5+1)}{A_{l_1}A_{l_2}A_{l_3}A_{l_4}A_{l_5}^2}\begin{pmatrix}
l_1 & l_2 & l_5\\
0 & 0 & 0
\end{pmatrix}^2\begin{pmatrix}
l_3 & l_4 & l_5\\
0 & 0 & 0
\end{pmatrix}^2\cr
&+  \frac{4}{A_0^2}\sum_{\boldsymbol{l}\neq \bold{1}}\frac{(2l_1+1)(2l_2+1)(2l_3+1)}{A_{l_1}^2 A_{l_2}A_{l_3}}~.
\end{align}
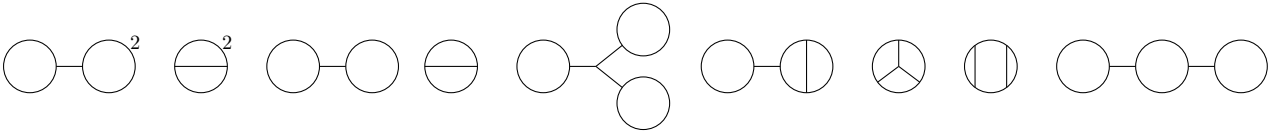
\begin{figure}[H]
\begin{center}
\begin{tikzpicture}[scale=.35]

\draw (-13.5,0) circle (1cm);
\draw[black] (-12.5,0) --(-11.5,0);
\draw (-10.5,0) circle (1cm);
\node[scale=.7] at (-9.5,.9) {2};

\draw (-7,0) circle (1cm);
\draw[black] (-8,0) --(-6,0);
\node[scale=.7] at (-6,.9) {2};

\draw (-3.5,0) circle (1cm);
\draw[black] (-2.5,0) --(-1.5,0);
\draw (-.5,0) circle (1cm);

\draw (2.5,0) circle (1cm);
\draw[black] (1.5,0) --(3.5,0);

\draw (6,0) circle (1cm);
\draw[black] (8,0) --(9,.8);
\draw[black] (8,0) --(9,-.8);
\draw[black] (8,0) --(7,0);
\draw (9.8,1.4) circle (1cm);
\draw (9.8,-1.4) circle (1cm);

\draw (13,0) circle (1cm);
\draw[black] (14,0) --(15,0);
\draw (16,0) circle (1cm);
\draw[black] (16,-1) --(16,1);

\draw (19.5,0) circle (1cm);
\draw[black] (19.5,0) --(19.5,1);
\draw[black] (19.5,0) --(18.7,-.6);
\draw[black] (19.5,0) --(20.3,-.6);
\draw (23.,0) circle (1cm);
\draw[black] (23.6,.8) --(23.6,-.8);
\draw[black] (22.4,.8) --(22.4,-.8);
\draw (26.5,0) circle (1cm);
\draw[black] (27.5,0) --(28.5,0);
\draw (29.5,0) circle (1cm);
\draw[black] (30.5,0) --(31.5,0);
\draw (32.5,0) circle (1cm);

\end{tikzpicture}
\end{center}
\caption{Diagrams appearing in the Wick contraction of $\varphi(\Omega)^3\varphi(\Omega')^3\varphi(\Omega'')^3\varphi(\Omega''')^3$.}
\end{figure}
\noindent
\subsection{Numerical evidence}
In this section we provide some numerical evaluations of the diagrams contributing to loops$_{\beta^4}$. 
By putting a sharp cutoff $\boldsymbol{L}= 250$ on the summation indices $\boldsymbol{l}$ of the melonic diagrams $\ominus$ (\ref{melons}) we test the conjecture (\ref{conjecture_melons})
\begin{equation}
\mathlarger{\mathlarger{\ominus}} -\frac{10}{3}+\frac{5\pi a_1}{2a_0}=0~,\quad \frac{a_1}{a_0}= \frac{27}{20\pi}~.
\end{equation}
We obtain $|\mathlarger{\mathlarger{\ominus}}| = 0.0411$ (while $1/24 \sim 0.0416$.). For two-loop$_{\beta^4}$, i.e. the last three contributions at order $\mathcal{O}(\beta^4)$ 
arising because of the FP contribution to the propagator we find 
\begin{align}
\text{two-loop}_{\beta^4} \sim 0.013\pi\times \frac{a_1}{a_0}~.
\end{align}
Putting sharp cutoffs on the other diagrams in loops$_\beta^4$ (\ref{3loop_diagrams}) we find convergence toward
\begin{align}
&\sum_{\boldsymbol{l}\neq \textbf{1}}^{\bold{200}} \frac{(2l_1+1)(2l_2+1)(2l_3+1)}{A_{l_1}^2A_{l_2} A_{l_3}} \begin{pmatrix} l_1 & l_2 & l_3 \\ 0 & 0 & 0\end{pmatrix}^2\sim  0.193~,\cr
&\sum_{\boldsymbol{l}\neq \bold{1}}^{\bold{60}}\sum_{l_5\geq 0}^{120}\frac{(2l_1+1)(2l_2+1)(2l_3+1)(2l_4+1)(2l_5+1)}{A_{l_1}A_{l_2}A_{l_3}A_{l_4}}\begin{pmatrix} l_1 & l_2 & l_5\\ 0 & 0 & 0\end{pmatrix}^2\begin{pmatrix} l_3 & l_4 & l_5\\ 0 & 0 & 0\end{pmatrix}^2\sim 2.615~,\cr
&\sum_{\boldsymbol{l}\neq \bold{1}}^{\bold{20}} \frac{(2l_1+1)(2l_2+1)(2l_3+1)(2l_4+1)(2l_5+1)}{A_{l_1}A_{l_2}A_{l_3}A_{l_4}A_{l_5}}\begin{pmatrix}  l_1 & l_2 & l_5 \\ 0 & 0 & 0\end{pmatrix}^2 \begin{pmatrix}  l_3 & l_4 & l_5 \\ 0 & 0 & 0\end{pmatrix}^2\sim -0.240~,\cr
&\sum_{\boldsymbol{l}\neq \bold{1}}^{\bold{30}}\frac{(2l_1+1)(2l_2+1)(2l_3+1)(2l_4+1)(2l_5+1)}{A_{l_1}A_{l_2}A_{l_3}A_{l_4}A_{l_5}^2}\begin{pmatrix}
l_1 & l_2 & l_5\\
0 & 0 & 0
\end{pmatrix}^2\begin{pmatrix}
l_3 & l_4 & l_5\\
0 & 0 & 0
\end{pmatrix}^2 \sim 0.125~.
\end{align}
The last diagram in loops$_{\beta^4}$ we could not evaluate for a high enough cutoff. It is negative and for some small cutoff $\boldsymbol{L}=\bold{4}$ is equal to $-0.048$, tending however to a larger negative value. Combining these results we obtain loops$_{\beta^4} - \mathlarger{\mathlarger{\ominus}}^2/2  \sim 1.28$.
Consequently in
\begin{align}
0&\overset{?}{=} -\frac{25\pi^2}{8}\left(\frac{a_1}{a_0}\right)^2+\frac{15\pi}{4}\left(\frac{a_1}{a_0}\right)- \frac{17}{27}+{\text{loops}}_{\beta^4}-\frac{1}{2}\mathlarger{\mathlarger{\ominus}}^2+\frac{2}{3}\zeta(3) ~,
\end{align}
we are off by around 0.6. We believe this is solely a numerical issue and could be solved by increasing the last cutoff.

 \begingroup

\end{document}